\documentclass[pra,reprint]{revtex4-1} 
\usepackage{amssymb,amsmath}
\usepackage{graphicx, import}
\usepackage{placeins}
\usepackage{hyperref}
\usepackage{color}

\newcommand{\state}[0]{f} 
\newcommand{\normpr}[0]{\tilde{L}_{BR} }
\newcommand{\ket}[1]{| #1 \rangle} 
\newcommand{\bra}[1]{\langle #1 |} 
\newcommand{\op}[1]{\hat{#1}} 
\newcommand{\p}[1]{\hat{\sigma}_{#1}} 
\newcommand{\idn}[0]{\op{\mathcal{I}}} 

\newcommand{\ex}[1]{\mathbb{E}[#1]} 
\newcommand{\norm}[1]{||#1||} 
\newcommand{\amx}[1]{\hat{x}_{#1}(-)} 
\newcommand{\apx}[1]{\hat{x}_{#1}(+)} 
\newcommand{\amp}[1]{\hat{P}_{#1}(-)} 
\newcommand{\app}[1]{\hat{P}_{#1}(+)} 

\graphicspath{ {./}} 

\begin{document}

\title{Machine learning for predictive estimation of qubit dynamics subject to dephasing}

\author{Riddhi Swaroop Gupta} 
\email{rgup9526@uni.sydney.edu.au}
\affiliation{ARC Centre of Excellence for Engineered Quantum Systems, School of Physics, The University of Sydney, New South Wales 2006, Australia}

\author{Michael J. Biercuk}
\affiliation{ARC Centre of Excellence for Engineered Quantum Systems, School of Physics, The University of Sydney, New South Wales 2006, Australia}

\begin{abstract}
Decoherence remains a major challenge in quantum computing hardware and a variety of physical-layer controls provide opportunities to mitigate the impact of this phenomenon through feedback and feedforward control. In this work, we compare a variety of machine learning algorithms derived from diverse fields for the task of state estimation (retrodiction) and forward prediction of future qubit state evolution for a single qubit subject to classical, non-Markovian dephasing. Our approaches involve the construction of a dynamical model capturing qubit dynamics via autoregressive or Fourier-type protocols using only a historical record of projective measurements.  A detailed comparison of achievable prediction horizons, model robustness, and measurement-noise-filtering capabilities for Kalman Filters (KF) and Gaussian Process Regression (GPR) algorithms is provided. We demonstrate superior performance from the autoregressive KF relative to Fourier-based KF approaches and focus on the role of filter optimization in achieving suitable performance. Finally, we examine several realizations of GPR using different kernels and discover that these approaches are generally not suitable for forward prediction.  We highlight the underlying failure mechanism in this application and identify ways in which the output of the algorithm may be misidentified numerical artefacts.
\end{abstract}

\maketitle

\section{Introduction} 

In predictive estimation, a dynamically evolving system is observed and any temporal correlations encoded in the observations are used to predict the future state of the system.  This generic problem is well studied in diverse fields such as engineering, econometrics, meteorology, and seismology~\cite{groen2013real,dong2009unscented,ko2009gp,harvey1990forecasting,cheng2015time}, and is addressed in the control-theoretic literature as a form of filtering.  Applying these approaches to state estimation on qubits is complicated by a variety of factors; dominant among these is the violation of the assumption of linearity inherent in most filtering applications as qubit states are formally bilinear. The case of an idling, or freely evolving qubit subject to dephasing is more complicated still, as an a priori model of system evolution suitable for implementation within standard filtering algorithms will not in general be available.

Fortunately there are many lessons to learn from classical control, even in the presence of such complications.  For classical systems, machine learning techniques have enabled state tracking, control, and forecasting for highly non-linear and noisy dynamical trajectories or complex measurement protocols (e.g. \cite{garcia2016optimal, bach2004learning, tatinati2013hybrid, hall2011reinforcement, hamilton2016ensemble}). These demonstrations move far beyond the simplified assumptions underlying many basic filtering tasks such as linear dynamics and white (uncorrelated) noise processes. For instance, so-called particle-based Bayesian frameworks (e.g. particle filtering, unscented or sigma-point filtering) allow state estimation and tracking in the presence of non-linearities in system dynamics or measurement protocols~\cite{candy2016bayesian}.  Further extensions approach the needs of a stochastically evolving system; recently, an ensemble of so-called unscented Kalman filters, named after the underlying mathematical transformation, demonstrated state estimation and forward predictions for chaotic, non-linear systems in the absence of a prescribed model~\cite{hamilton2016ensemble}. For non-chaotic, multi-component stationary random signals, other algorithmic approaches have been particularly useful for tracking instantaneous frequency and phase information, ~\cite{boashash1992estimating2, ji2016gradient}, enabling short-run forecasting.  

In the field of quantum control, work has begun to incorporate the additional challenges faced when considering state estimation on qubits, notably quantum-state collapse under projective measurement.  Under such circumstances, in which the measurement backaction strongly influences the quantum state (in contrast with the classical case), it is not straightforward to extend machine learning predictive estimation techniques.  Work to date has approached the analysis of projective measurement records on qubits as pattern recognition or image reconstruction problems, for example, in characterising the initial or final state of quantum system (e.g. \cite{struchalin2016experimental, sergeevich2011characterization, mahler2013adaptive}) or reconstructing the historical evolution of a quantum system based on large measurement records (e.g. \cite{stenberg2016characterization, shabani2011efficient, shen2014reconstructing, de2016estimation, tan2015prediction, huang2017neural}). In adaptive or sequential Bayesian learning applications, a projective measurement protocol may be designed or adaptively manipulated to efficiently yield noise-filtered information about a quantum system (e.g. \cite{bonato2016optimized, wiebe2015bayesian,shulman2014suppressing,granade2016practical}). 

The demonstrations above typically assume the object of interest is either static, or stochastically evolves in a manner which is dynamically uncorrelated in time (white) as measurement protocols are repeated. This simplifying assumption falls well short of typical laboratory based experiments where noise processes are frequently correlated in time, and evolution may also occur rapidly relative to a measurement protocol. In such a circumstance, further complexity is introduced as the Markov condition commonly assumed in Bayesian learning frameworks~\cite{candy2016bayesian} is immediately violated.  Even in the classical case, the problem of designing an appropriate representation of non-Markovian dynamics in Bayesian learning frameworks is an active area of research (e.g  \cite{jacob2017bayesian}).  Hence, the canonical real-time tracking and prediction problem - where a non-linear, stochastic trajectory of a system is tracked using noisy measurements and short-run forecasts are made - is under-explored for quantum systems with projective measurements.

In this manuscript, we develop and explore a broad class of predictive estimation algorithms allowing us to track a qubit state undergoing \emph{stochastic but temporally correlated} evolution using a record of projective measurements, and forecast its future evolution. Our approaches employ machine learning algorithms to extract temporal correlations from the measurement record and use this information to build an effective dynamical model of the system's evolution.  We design a deterministic protocol to correlate Markovian processes such that a certain general class of non-Markovian dynamics can be approximately tracked without violating the assumptions of a machine learning protocol, based on the theoretically accessible and computationally efficient frameworks of Kalman Filtering (KF) and Gaussian Process Regression (GPR).  Both frameworks provide a mechanism by which temporal correlations (equally, dynamics) are encoded into an algorithm's structure such that projection of data-sets onto this structure enables meaningful learning, white-noise filtering, and effective forward prediction.  We perform numerical simulations to test the effectiveness of these algorithms in maximizing the prediction horizon under various conditions, and quantify the role of the measurement sampling rate relative to the noise dynamics in defining the prediction horizon.  Simulations incorporate a variety of measurement models, including pre-processed data yielding a continuous measurement outcome and discretised outcomes commonly associated with single-shot projective qubit measurements.   We find that in most circumstances an autoregressive Kalman framework yields the best performance, providing model-robust forward prediction horizons and effective filtering of measurement noise.  Finally, we demonstrate that standard GPR-based protocols employing a variety of kernels, while effective for the problem of filtering (fitting) a measurement record, are not suitable for real-time forecasting beyond the measurement record.  

In what follows, we describe in detail the physical setting for our problem in~Section~\ref{sec:main:PhysicalSetting} and explain how this leads to a specific choice of algorithm which may be deployed for the task of tracking non-Markovian state dynamics in the absence of a dynamical model for system evolution.  We provide an overview of the central GPR and KF frameworks in Section~\ref{sec:main:OverviewofPredictive Methodologies}, and we specify a series of algorithms under consideration in this paper tailored to different measurement processes. For pre-processed measurement records, we consider four algorithmic approaches: a Least Squares Filter (LSF) from \cite{mavadia2017}; an Autoregressive Kalman Filter (AKF); a so-called Liska Kalman Filter from \cite{livska2007} adapted for a Fixed oscillator Basis (LKFFB); and a suitably designed GPR learning protocol. For binary measurement outcomes, we extend the AKF to a Quantised Kalman Filter (QKF). In Section~\ref{sec:main:Optimisation}, we present optimisation procedures for tuning all algorithms. Numerical investigations of algorithmic performance are presented in Section~\ref{sec:main:Performance} and a comparative analysis of all algorithms is provided in Section~\ref{sec:main:discussion}. 

\section{Physical Setting \label{sec:main:PhysicalSetting}}  
\label{sec:main:1} 

Our physical setting considers a sequence of projective measurements performed on a qubit. Each projective measurement yields a 0 or 1 outcome representing the state of the qubit. The qubit is then reset, and the exact procedure is repeated. By considering a qubit state initialized in a superposition of the measurement basis (for us, Pauli $\p{z}$ eigenstates), we gain access to a direct probe of qubit phase evolution.  If, for instance, no dephasing is present, then the probability of obtaining a binary outcome remains static in time as sequential qubit measurements are performed. If slowly drifting environmental dephasing is present, then the probability of obtaining a given binary outcome also drifts stochastically. In essence, the qubit probes dephasing noise and our procedure encodes a continuous-time non-Markovian dephasing process into time-stamped, discrete binary samples through the nonlinear projective measurement, carrying the underlying correlations in the noise.  It is this series of measurements which we seek to process in our algorithmic approaches to qubit state tracking and prediction. 

\begin{figure}[h!]
	\includegraphics[scale=1]{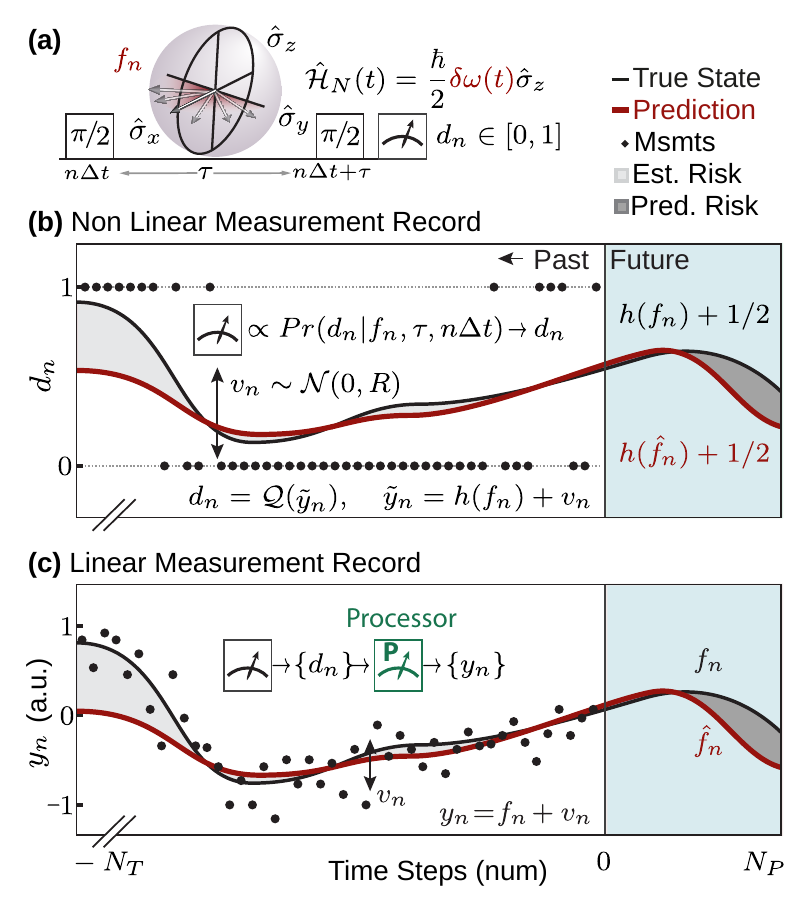} 
	\caption{ \label{fig:main:Predive_control_Fig_overview_17_one} (a) A Ramsey experiment at $t=n\Delta t$ with fixed wait time $\tau$ and time-steps, $n$, spaced $\Delta t > \tau$ apart. A $\pi/2$ pulse rotates qubit state to super-position of $\ket{d}$ states, $d\in \{0,1\}$; qubit evolves via $\op{\mathcal{H}}_N(t)$ accumulating relative stochastic $\state_n$, for non-zero environmental dephasing $\delta \omega (t)$. Jittering arrows depict potential qubit state vectors permitted for (unknown) random $f_n$. Qubit state is measured as $d_n=d$ in $\p{z}$ basis after a second $\pi/2$ rotation. (b) Black dots depict $\{d_n\}$ against time steps, $n$; data collection stops at $n=0$ separating past state estimation from future prediction [blue region].  Black solid line shows true qubit state likelihood $ \propto h(f_n)$; and  red solid line shows state estimate (prediction) for $n<0$ ($n>0$). A prediction horizon is $n < n^* \in [0,N_P]$ for which dark-grey region between red and black lines is minimised (Bayes prediction risk) relative to predicting the mean of dephasing noise; algorithmic tuning occurs by minimising light-grey region (Bayes state estimation risk). $\mathcal{Q}$ quantises black line into noisy qubit measurements, $d_n$, under Gaussian uncertainty $v_n$. (c) Single shot outcomes in (b) are pre-processed to yield noisy measurements $\{ y_n\}$ [black dots]; $y_n$ is linear in $\state_n$ and $v_n$ represents additive white Gaussian measurement noise.}
\end{figure}

Formally, an arbitrary environmental dephasing process manifests as time-dependent stochastic detuning, $\delta \omega (t)$, between the qubit frequency and the system master clock. This detuning is an experimentally measurable quantity in a Ramsey protocol, as shown schematically in Fig.~\ref{fig:main:Predive_control_Fig_overview_17_one} (a). A non-zero detuning over measurement period $\tau$ (starting from $t=0$) induces a stochastic relative phase accumulation (in the rotating frame) for a qubit superposition state as $\left|0\right\rangle+e^{-i\state(0, \tau)}\left|1\right\rangle$ between qubit basis states.  The accumulated $\state(0, \tau)$ at the end of a single Ramsey experiment is mapped to a probability of obtaining a particular outcome in the measurement basis via the form of the Ramsey sequence.  

In a sequence of $n$ Ramsey measurements spaced $\Delta t$ apart with a fixed duration, $\tau$, the change in the statistics of measured outcomes over this measurement record depends solely on the dephasing  $\delta \omega(t)$.   We assume that the measurement action over $\tau$ is much faster than the temporal dynamics of the dephasing process, and $\Delta t \gtrsim \tau$. The resulting measurement record is a set of binary outcomes,  $\{d_n\}$, determined probabilistically from $n$ true stochastic qubit phases, $\state := \{\state_n\}$. Here the accumulated phase in each Ramsey experiment, $ \state(n \Delta t, n\Delta t + \tau) \equiv \int_{n \Delta t}^{n \Delta t +\tau} \delta \omega(t') dt'$ and we use the shorthand $\state(n \Delta t , n\Delta t + \tau) \equiv \state_n$.  We define the statistical likelihood for observing a single shot, $d_n$, using Born's rule \cite{ferrie2013}:

\begin{align}
	Pr(d_n=d | \state_n, \tau, n \Delta t) &= \begin{cases} \cos^2(\frac{\state_{n}}{2}) \quad \text{for $d=1$} \\   \sin^2(\frac{\state_{n}}{2})  \quad \text{for $ d=0$}  \end{cases} \label{eqn:main:likelihood} 
\end{align}
The notation $Pr(d_n | \state_n, \tau, n \Delta t)$ refers to the conditional probability of obtaining measurement outcome $d_n$ given a true stochastic phase, $\state_n$, accumulated over $\tau$, beginning at time $t = n \Delta t$. In the noiseless case, $Pr(d_n=1|\state_n, \tau, n \Delta t) = 1, \quad \forall n $, such that a qubit exhibits no additional phase accumulation due to environmental dephasing. Following a single measurement the qubit state is reset, but the dephasing noise correlations manifest again via Born's rule for another random value of the bias at time-step $n+1$. A detailed discussion of Eq.~(\ref {eqn:main:likelihood}) can be found in Appendix~\ref{sec:app:setup_1}.

The action of measurement, expressed as $h(\state_n)$, is given by $Pr(d_n=d| \state_n, \tau, n \Delta t) \equiv \frac{1}{2} - (-1)^d h(\state_n) $ and is depicted in Fig.~\ref{fig:main:Predive_control_Fig_overview_17_one}(b) as a probability of seeing the qubit in the $d=1$ state.  We begin by describing here a `raw' non-linear measurement record, $\{ d_n\}$ where each $d_n$ [black dots] corresponds to a binary outcome derived from a single projective measurement on a qubit. The sequence $\{ d_n\}$ can be treated as a sequence of biased coin flips, where the underlying bias of the coin is a non-Markovian, discrete-time process and the value of the bias is given by Eq.~(\ref {eqn:main:likelihood}) at each $n$. The non-linearity of the measurement, $h(\state_n)$, is defined with respect to $\state_n$ where Eq.~(\ref {eqn:main:likelihood}) is interpreted as a non-linear measurement action for Bayesian learning frameworks.

This data series is contrasted with a linear measurement record, $\{ y_n\}$, depicted in Fig.~\ref{fig:main:Predive_control_Fig_overview_17_one}(c).  Each value $y_n$ is derived from the sum of a true qubit phase, $\state_n$, and Gaussian white measurement noise, $v_n$.  The sequence $\{ y_n\}$ is generated by pre-processing raw binary measurements, $\{ d_n\}$ via a range of experimental techniques subject to a separation of timescales such that $\sim\tau$ is much faster than drift of $\delta \omega (t)$.  In the most common case, one performs $M$ runs of the experiment over which $\delta \omega (t)$ is approximately constant, giving an estimate of  $\state_n$ at $t = n \Delta t $ using averaging, a Bayesian scheme, or Fourier analysis. A more complex linearization protocol involves the use of low-pass or decimation filtering on a sequence $\{ d_n\}$  to yield $\hat{Pr}(d_n | \state_n, \tau, n\Delta t)$, from which accumulated phase corrupted by measurement noise, $\{ y_n\}$, can be obtained from Eq.~(\ref {eqn:main:likelihood}). 

We impose properties on environmental dephasing such that our theoretical designs can enable meaningful predictions. We assume dephasing is non-Markovian, covariance stationary and mean-square ergodic.  That is, a single realisation of the process $\state$ is drawn from a power spectral density of arbitrary, but non-Markovian form. We further assume that $\state$ is a Gaussian process and the separation of timescales between measurement protocols and dephasing dynamics articulated above are met.

Given these conditions, our task is to build a dynamical model to approximately track $\state$ over past measurements ($n<0$), and enable qubit state predictions in future times ($n>0$).  This prediction is represented by the red line in Fig.~\ref{fig:main:Predive_control_Fig_overview_17_one}(b-c), and differs from the truth by the so-called estimation (prediction) risk for past (future) times as indicated by shading.  We represent our estimate of $\state$ for all times using a hat in both the linear and nonlinear measurement models.  The major challenge we face in developing this estimate, $\hat{\state}$ (equivalently $\hat{Pr}(d_n | \state_n, \tau, n\Delta t)$), is that for a qubit evolving under stochastic dephasing (true state given by black solid line in Fig.~\ref{fig:main:Predive_control_Fig_overview_17_one}(b) and (c)), we have no a prior dynamical model for the underlying evolution of $\state$.  In the next section, we define the theoretical structure of KF and GPR algorithms which allow us to build that dynamical model directly from the historical measurement record.

\section{Overview of Predictive Methodologies \label{sec:main:OverviewofPredictive Methodologies}}

\begin{figure*}[htp]
	\includegraphics[scale=1.]{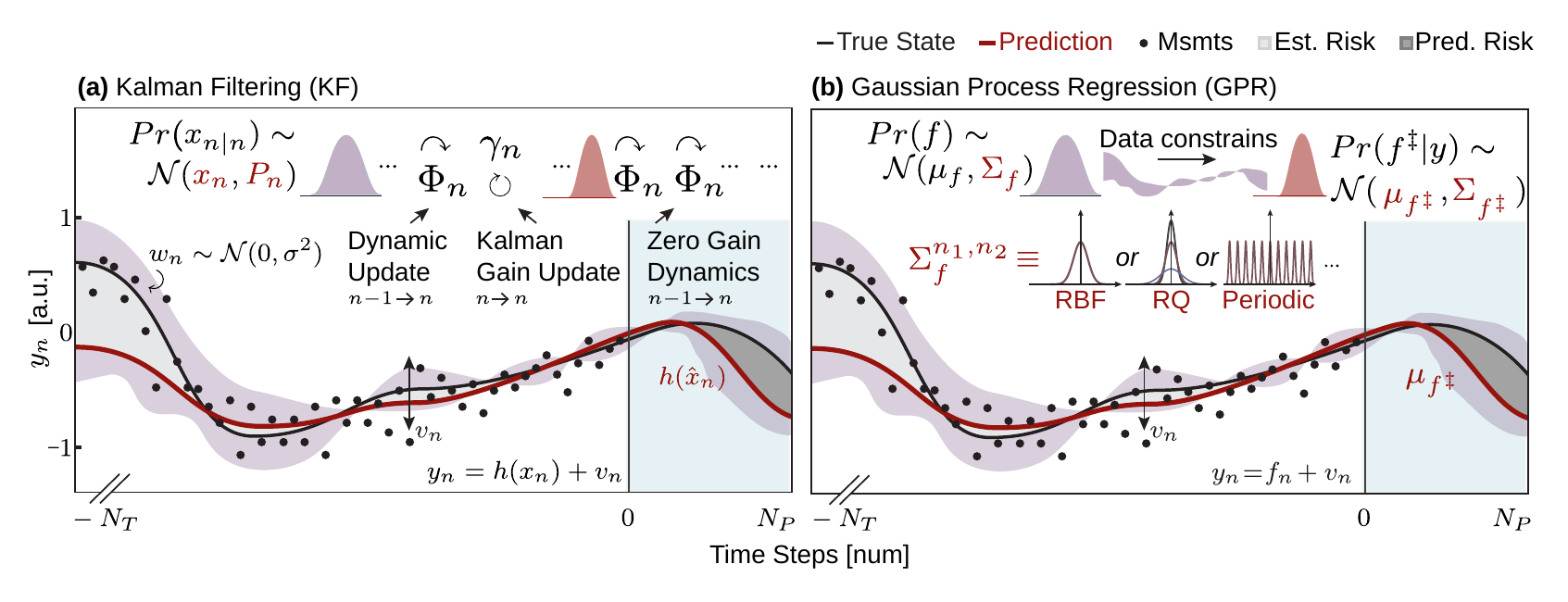} 
	\caption{ \label{fig:main:Predive_control_Fig_overview_17_two} Comparison of the algorithmic structure between the KF and GPR by superposing lower panels of Fig.~\ref{fig:main:Predive_control_Fig_overview_17_one} with KF and GPR predictive frameworks. (a) KF: Purple distribution represents a prior, with mean $x_n$, and covariance $P_n$; propagated in time-steps, $n$, using Kalman dynamics $\Phi_n$, and updated within each $n$ by the Kalman gain $\gamma_n$ to yield  posterior distribution (red) at $n$. The posterior at $n$ is the prior at $n+1$. The mean of a posterior distribution at each $n$ is used to derive predictions given by the red line using $h(x_n)$. In blue region, the red posterior predictive distribution is propagated using $\Phi_n$ but $\gamma_n \equiv 0$. Gaussian white Kalman `process' noise, $w_n$, is coloured by $\Phi_n$ to yield dynamics for $x_n$. (b) Purple prior distribution defined over sequences, $\state$, with mean, $\mu_\state$ and variance $\Sigma_\state$ is constrained by the entire measurement record. The resulting posteriori predictive distribution (red) is evaluated at test-points in time, $n^\ddagger \in [-N_T, N_P]$; state estimates (predictions) is the mean, $\mu_{\state^\ddagger}$ at $n^\ddagger < 0$ ($n^\ddagger > 0$). A choice of kernel defines each element in $\Sigma_{\state}, \Sigma_{\state^\dagger}$. In both (a)-(b), the purple shadow represents posterior state variance (diagonal $P_n$ or $\Sigma_{\state^\ddagger}$ elements) constrained by data and filtered measurement noise $v_n$.}
\end{figure*}

Our objective is to implement an algorithm permitting learning of underlying qubit dynamics in such a way as to maximize the forward prediction horizon for a given qubit data record.  We first quantify the quality of our state estimation procedure.  The fidelity of any underlying algorithm during state estimation and prediction, relative to the true state, is expressed by the mathematical quantity known as a Bayes Risk, where zero risk corresponds to perfect estimation. At each time-step, $n$, the Bayes risk is a mean square distance between truth, $\state$, and prediction, $\hat{\state}$, calculated over an ensemble of $M$ different realisations of true $\state$ and noisy data-sets $\mathcal{D}$:
\begin{align}
	L_{BR}(n | I) & \equiv \langle(\state_n - \hat{\state}_n)^2 \rangle_{\state,\mathcal{D}} \label{eqn:main:sec:ap_opt_LossBR}
\end{align}
The notation $L_{BR}(n | I)$ expresses that the Bayes Risk value at $n$ is conditioned on $I$, a placeholder for free parameters in the design of the predictor, $\hat{\state}_n$. State estimation risk is Bayes Risk incurred during $n \in [-N_T, 0]$; prediction risk is the Bayes Risk incurred during $n \in [0, N_P]$. State estimation and prediction risk regions for one realisation of dephasing noise are shaded in Fig.~\ref{fig:main:Predive_control_Fig_overview_17_one}-\ref{Predive_control_Fig_overview_17_three}.  We therefore define the forward prediction horizon as the number of time-steps for $ n^{*} \in [0, N_P]$ during which a predictive algorithm incurs a lower Bayes prediction risk than naively predicting $\hat{\state}_n \equiv \mu_f = 0 \quad \forall n$, the mean qubit behaviour under zero-mean dephasing noise. 

With this concept in mind, we introduce two general approaches for algorithmic learning relevant to the strictures of the problem we have introduced.  Our general approach is shared between all algorithms employed and is represented schematically for the KF and GPR in Fig.~\ref{fig:main:Predive_control_Fig_overview_17_two}. Stochastic qubit evolution is depicted for one realisation of $\state$ [black solid line] given noisy linear measurements [black dots] corrupted by Gaussian white measurement noise $v_n$.  Our overall task is to produce an estimate, given by the red line, which minimizes risk for the prediction period.  Ideally both estimation risk and prediction risk are minimized simultaneously for well performing implementations.

Examining the insets in both panels of Fig.~\ref{fig:main:Predive_control_Fig_overview_17_two}, both frameworks start with a prior Gaussian distribution over qubit states [purple] that is constrained by the measurement record to yield a posterior Gaussian distribution of the qubit state [red]. The prior captures assumptions about the qubit state before any data is seen and the posterior expresses our best knowledge of the qubit state under a Bayesian framework.  The posterior distribution in both KF and GPR is used to generate qubit state estimates ($n<0$) and predictions ($n>0$) [red solid line].  However the computational process by which this posterior is inferred differs significantly between the two methods; we provide an overview of the central features of these algorithms below. 

The key feature of a Kalman filter is the recursive learning procedure shown in the inset to Fig.~\ref{fig:main:Predive_control_Fig_overview_17_two}(a). Our knowledge of the qubit state is summarised by the prior and a posterior Gaussian probability distributions and these are created and collapsed recursively \emph{at each time step}. The mean of these distributions is the true Kalman state, $x_n$, and the covariance of these distributions, $P_n$, captures the uncertainty in our knowledge of $x_n$; together both  define the Gaussian distribution. The Kalman filter produces an \emph{estimate} of the state, $\hat{x}_{n}$ at each step through this recursive procedure taking into account two factors. First, the Kalman gain, $\gamma_n$, updates our knowledge of $(x_n, P_n)$ within each time step $n$ and serves as a weighting factor for the difference between  incoming data, and our best estimate for an observation based on $\hat{x}_n$, suitably transformed via the measurement action, $h(\hat{x}_{n})$. Next, the dynamical model $\Phi_n$ propagates the state and covariance, $(x_n, P_n)$, to the next time step, such that the posterior moments at $n$ define the prior at $n+1$.  This process occurs for each time step and an estimate of a true $x_n$ state is built up recursively based on all of our existing knowledge, namely, a linear combination of all past measurements; and all previously generated state estimates.  Beyond $n=0$ we perform predictions in the absence of further measurement data by simply propagating the dynamic model with the Kalman gain set to zero.  Full details of the KF algorithm appear below in Section~\ref{Subsec:KF}.

In our application, we define the Kalman state, $x_n$, the dynamical model $\Phi_n$, and a measurement action $h(x_n)$ such that the Kalman Filtering framework can track a non-Markovian qubit state trajectory due to an arbitrary realisation of $\state$. In standard KF implementations, the discrete-time sequence $\{x_n\}$, defines a ``hidden'' signal that cannot be observed, and the dynamic model $\Phi_n$ is known.  We deviate from this standard construction such that our true Kalman state and its uncertainty, $(x_n, P_n)$, do not have a direct physical interpretation.  Kalman $x_n$ has no a priori deterministic component and corresponds to arbitrary power spectral densities describing $\state$. Hence, the role of the Kalman $x_n$ is to represent an abstract correlated process that, upon measurement, yields physically relevant quantities governing qubit dynamics.  Moreover a key challenge described in detail below is to construct an effective $\Phi_{n}$ from the measurement record.   

In contrast to the recursive approach taken in the KF, a GPR learning protocol illustrated schematically in Fig.~\ref{fig:main:Predive_control_Fig_overview_17_two}(b) selects \textit{a random process} to best describe overall dynamical behaviour of the qubit state under one realisation of $\state$. The key point is that sampling the prior or posterior distribution in GPR yields random realisations of discrete time \textit{sequences}, rather than individual random variables, and GPR considers the entire measurement record at once.  In a sense, it corresponds to a form of fitting over the entire data set.  The output of a GPR protocol is a predictive distribution which we can evaluate at an arbitrarily chosen sequence of test-points, where the test points can exist  for $n<0$ ($n>0$) such that we extract state estimates (forward predictions) from the predictive distribution. Due to the nature of this procedure, we wish to distinguish the set of test points (in units of time-steps)  using a ${}^\ddagger$, namely, that we are evaluating the predictive posterior distribution of a GPR protocol at desired time labels. In this notation, $\{ n^{\ddagger} \}, \quad n^{\ddagger} \in [-N_T, N_P]$ are test-points; $N^{\ddagger}$ is the total length of an array of test points; where state estimation occurs if $n^{\ddagger} \leq 0$ and predictions occur if $n^{\ddagger}>0$. 

The process of building the posterior distribution is implemented using a kernel, or basis, from which to construct the effective fit.  In standard GPR implementations, the correlation between any two observations depends only on the separation distance of the index of these observations, and correlations are captured in the covariance matrix, $\Sigma_\state$. Each element, $\Sigma_\state^{n_1, n_2}$, describes this correlation for observations at arbitrary time-steps indexed by $n_1$ and $n_2$: this quantity is given in a form set by the selected kernel. 

In our application, the non-Markovian dynamics of $\state$ are not specified explicitly but are encoded in a general way through the choice of kernel, prescribing how $\Sigma_\state^{n_1, n_2}$ should be calculated. The Fourier transform of the kernel represents a power spectral density in Fourier space. A general design of $\Sigma_\state^{n_1, n_2}$ allows one to probe arbitrary stochastic dynamics and equivalently, explore arbitrary regions in the Fourier domain. For example, Gaussian kernels (RBF) and mixtures of Gaussian kernels (RQ) capture the continuity assumption that correlations die out as separation in time increases. We choose to employ an infinite basis of oscillators implemented by the so-called periodic kernel to enable us to represent arbitrary power spectral densities for $\state$.  Prediction occurs simply by extending the GPR fit by choosing test-points $n^{\ddagger}>0$.

In the following subsections we provide details of the specific classes of learning algorithm employed here with an eye towards evaluating their predictive performance on qubit-measurement records.  We introduce a series of KF algorithms capable of handling both linear and non-linear measurement records, and restrict our analysis of GPR to linear measurement records.

\subsection{ Kalman Filtering (KF)}\label{Subsec:KF}

In order for a Kalman Filter to track a stochastically evolving qubit state in our application, the hidden true Kalman state at time-step $n$, $x_n$, must mimic stochastic dynamics of a qubit under environmental dephasing. We propagate the hidden state $x_n$ according to a dynamical model $\Phi_n$ corrupted by Gaussian white process noise, $w_n$.  
\begin{align}
	x_n & = \Phi_n x_{n-1} + \Gamma_n w_n \label{eqn:KF:dynamics} \\
	w_n & \sim \mathcal{N}(0, \sigma^2) \quad \forall n 
\end{align}
Process noise has no physical meaning in our application - $w_n$ is shaped by $\Gamma_n$ and deterministically colored by the dynamical model $\Phi_n$ to yield a non-Markovian $x_n$ representing qubit dynamics under generalised environmental dephasing. In addition to coloring via the dynamical model, the process noise covariance matrix, $Q_n \equiv \Gamma_n\Gamma_n^T $, offers an additional mechanism to shape input white noise by designing $\Gamma_n$.

We measure $x_n$ using an ideal measurement protocol, $h(x_n)$, and incur additional Gaussian white measurement noise $v_n$ with scalar covariance strength $R$, yielding scalar noisy observations $y_n$:
\begin{align}
	y_n &= z_n + v_n \\
	z_n & \equiv  h(x_n) \\
	v_n & \sim \mathcal{N}(0, R) \quad \forall n
\end{align}
The measurement procedure, $h(x_n)$, can be linear or non-linear, allowing us to explore both regimes in our physical application.

With appropriate definitions, the Kalman equations below specify all Kalman algorithms in this paper. At each time step, $n$, we denote estimates of the moments of the prior and posterior distributions (equivalently, estimates of the true Kalman state) with $(\amx{n}, \amp{n})$ and $(\apx{n}, \app{n})$ respectively. The Kalman update equations take a generic form (c.f.~\cite{grewal2001theory}) :

\begin{align}
	\amx{n} & = \Phi_{n-1} \apx{n-1} \label{eqn:main:KF:dynamic_x}\\ 
	Q_{n-1} & = \sigma^2 \Gamma_{n-1}\Gamma_{n-1}^T  \label{eqn:main:KF:Q}\\
	\amp{n}&= \Phi_{n-1} \app{n-1} \Phi_{n-1}^T + Q_{n-1} \label{eqn:main:KF:dynamic_P}\\
	\gamma_n &= \amp{n} H_n^T(H_n\amp{n}H_n^T + R_n)^{-1} \label{eqn:main:KF:gain}\\
	\hat{y}_n(-) & = h(\amx{n}) \label{eqn:main:KF:step_ahead}\\
	\apx{n} &= \amx{n} + \gamma_n (y_n - \hat{y}_n(-)) \label{eqn:main:KF:bayesian_x}\\
	\app{n} &= \left[1  - \gamma_n H_n \right] \amp{n} \label{eqn:main:KF:bayesian_p}
\end{align}
To reiterate, Eq.~(\ref {eqn:main:KF:dynamic_x}) and Eq.~(\ref {eqn:main:KF:dynamic_P}) bring the best state of knowledge from the previous time step into the current time step, $n$, as a prior distribution. Dynamical evolution is modified by features of process noise, as encoded in Eq.~(\ref {eqn:main:KF:Q}), and propagated in Eq.~(\ref {eqn:main:KF:dynamic_P}). The propagation of the moments of the a priori distribution, as outlined thus far, does not depend on the incoming measurement, $y_n$, but is determined entirely by the a priori (known) dynamical model, in our case $\Phi \equiv \Phi_n, \forall n$. 

The Kalman gain in Eq.~(\ref {eqn:main:KF:gain}) depends on the uncertainty in the true state, $\amp{n}$ and is modified by features of the measurement model, $H_n$, and measurement noise, $R_n \equiv R,\; \forall n$. It serves as an effective weighting function for each incoming observation.  Before seeing any new measurement data, the filter predicts an observation $\hat{y}_n(-)$ corresponding to the best available knowledge at $n$ in Eq.~(\ref {eqn:main:KF:step_ahead}). This value is compared to the actual noisy measurement $y_n$ received at $n$, and the difference is used to update our knowledge of the true state via Eq.~(\ref {eqn:main:KF:bayesian_x}). If measurement data is noisy and unreliable (high $R$), then $\gamma$ has a small value, and the algorithm propagates Kalman state estimates according to the dynamical model and effectively ignores data. In particular, only the second terms in both Eq.~(\ref {eqn:main:KF:bayesian_x}) and Eq.~(\ref {eqn:main:KF:bayesian_p}) represent the Bayesian update of the moments of a prior distribution ($(-)$ terms) to the posterior distribution ($(+)$ terms) at $n$. If $\gamma_n \equiv 0$, then the prior and posterior moments at any time step are exactly identical by Eqs.~(\ref {eqn:main:KF:bayesian_x}) and (\ref {eqn:main:KF:bayesian_p}), and only dynamical evolution occurs using Eqs.~(\ref {eqn:main:KF:dynamic_x}) to (\ref {eqn:main:KF:dynamic_P}).  This is the condition we employ when we seek to make forward predictions beyond a single time-step, and hence we set $\gamma \equiv 0$ during future prediction.

Since we do not have a known dynamical model $\Phi$ for describing stochastic qubit dynamics under $\state$, we will need to make design choices for  $\{ x, \Phi, h(x), \Gamma \}$  such that $\state$ can be approximately tracked. These design choices will completely specify algorithms introduced below and represent key findings with respect to our work in this manuscript. For a linear measurement record, $h(x) \mapsto Hx$ and we compare predictive performance for $\Phi$ modeling stochastic dynamics either via so-called `autoregressive' processes in the AKF, or via projection onto a collection of oscillators in the LKKFB.  In addition, we use the dynamics of AKF to define a Quantised Kalman filter (QKF) with a non-linear, quantised measurement model such that the filter can act directly on binary qubit outcomes. We provide the relevant details in sub-sections below.

\subsubsection{Autoregressive Kalman Filter (AKF)}

Recursive autoregressive methods are well-studied in classical control applications  (\emph{c.f.}~\cite{moon2006real}) presenting opportunities to leverage existing engineering knowledge in developing quantum control strategies.  In our application, we use an autoregressive Kalman filter to probe arbitrary, covariance-stationary qubit dynamics such that the dynamic model is constructed as a weighted sum of $q$ past values driven by white noise {\em i.e.} an autoregressive process of order $q$, AR($q$). Using Wold's decomposition, it can be shown that any zero mean covariance stationary process representing qubit dynamics has a representation in the mean-square limit by an autoregressive process of finite order, as in Appendix~\ref{sec:app:AKF}.

The study of AR($q$) processes falls under a general class of techniques based on autoregressive moving average (ARMA) models in adaptive control engineering and econometrics (e.g.  \cite{landau1998adaptive,hamilton1994time} respectively). For high-$q$ models in a typical time-series analysis, it is possible to decompose an AR($q$) into an ARMA model with a small number of parameters \cite{brockwell1996introduction, salzmann1991detection}. However, we retain a high-$q$ model to probe arbitrary power spectral densities. Further, literature suggests employing a high-$q$ model is relatively easier than a full ARMA estimation problem and enables lower prediction errors \cite{wahlberg1989estimation,brockwell1996introduction}. 

To construct the Kalman dynamical operator $\Phi$ for the AKF, we introduce a
set of $q$ coefficients $\{\phi_{q' \leq q}\}, q' = 1, ... , q $ to specify the dynamical model:
\begin{align}
	\state_n &= \phi_1 \state_{n-1} + \phi_2 \state_{n-2} + ... + \phi_q \state_{n-q} + w_n \label{eqn:main:ARprocess}
\end{align}

\begin{figure} [tp]
	\includegraphics[scale=1]{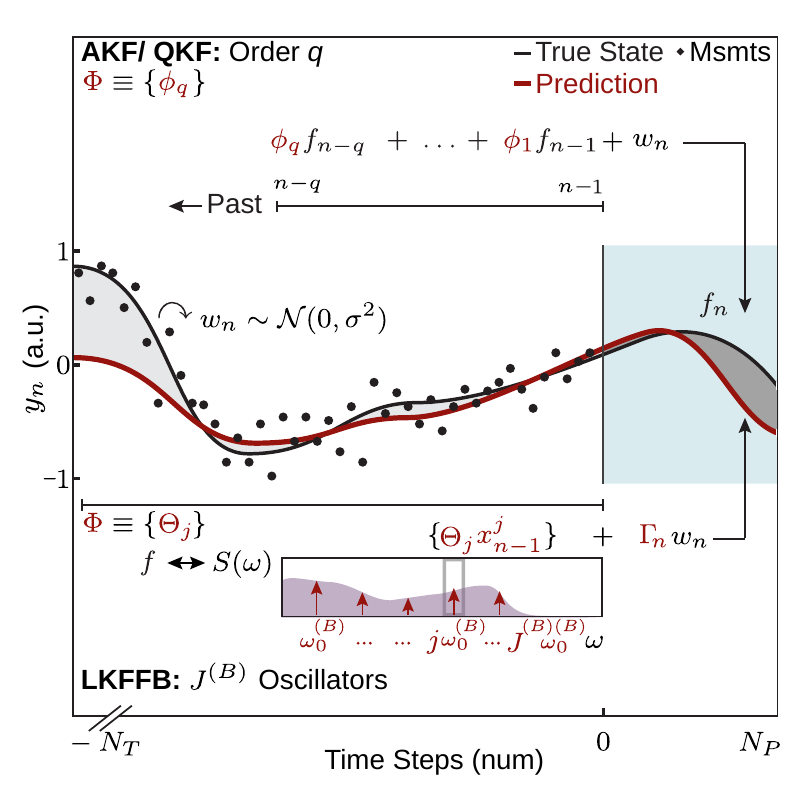}
	\caption{\label{Predive_control_Fig_overview_17_three} Approaches to construction of the KF dynamical model.  Panel (a) from Fig.~\ref{fig:main:Predive_control_Fig_overview_17_two} superimposed with Kalman dynamical models, $\Phi \equiv \Phi_n, \forall n$.  (a) AKF/QKF: A set of autoregressive coefficients, $\{\phi_{q'\leq q}\}$, define $\Phi$ to yield $f_n$ as a weight sum of $q$ past measurements. (b) LKFFB: Red arrows with heights $\norm{x^j_n}$ depict set of basis oscillators for $j = 1, \hdots, J^{(B)}$ probe true purple spectrum of $\state_n$ and yields time domain dynamics of $\state_n$ as a stacked system of resonators, $\Theta_j$. Black L-shaped arrows depict a single instance of $\state_n$ at $n=0$ based on historical $\{f_{n-1}, f_{n-2}, \hdots\} $.} 	
\end{figure}

\noindent We thus see that the dynamical model is constructed as a weighted sum of time-retarded samples of $\state$, with weighting factors given by the autoregressive coefficients up to order (and hence time lag) $q$. For small $q < 3$, it is possible to extract simple conditions on the coefficients, $\{ \phi_{q' \leq q} \}$, that guarantee properties of $\state$: for example, that $\state$ is covariance stationary and mean square ergodic. In our application, we freely employ arbitrary-$q$ models via machine learning in order to improve our approximation of an arbitrary $\state$. Any AR($q$) process can be recast (non-uniquely) into state space form (\cite{harvey1990forecasting}), and we define the AKF by the following substitutions into Kalman equations:
\begin{align}
	x_n & \equiv  \begin{bmatrix} f_{n} \hdots f_{n-q+1} \end{bmatrix}^T \\
	\Gamma_n w_n & \equiv \begin{bmatrix} w_{n} 0 \hdots 0 \end{bmatrix}^T \\
	\Phi_{AKF} & \equiv 
	\begin{bmatrix}
		\phi_1 & \phi_2 & \hdots & \phi_{q-1} & \phi_q \\ 
		1 & 0 & \hdots & 0 & 0 \\  
		0 & 1 & \ddots & \vdots & \vdots \\ 
		0 & 0 & \ddots & 0 & 0 \\ 
		0 & 0 & \hdots & 1 & 0 
	\end{bmatrix} \quad \forall n \label{eqn:akf_Phi} \\
	H & \equiv \begin{bmatrix} 1\;\;0\;\;0\;\;0\hdots0 \end{bmatrix} \quad \forall n  
\end{align}
The matrix $\Phi_{AKF}$ is the dynamical model used to recursively propagate the unknown state during state estimation in the AKF, as represented schematically in the upper half of Fig.~\ref{Predive_control_Fig_overview_17_three}. In general, the $\{\phi_{q' \leq q}\}$ employed in $\Phi_{AKF}$ must be learned through an optimisation procedure using the measurement record, where the set of parameters to be optimised is $\{\phi_1, \hdots, \phi_q, \sigma^2, R \}$. This procedure yields the optimal configuration of the autoregressive Kalman filter, but at the computational cost of a $q+2$-dimensional Bayesian learning problem for arbitrarily large $q$.

The Least Squares Filter (LSF) in \cite{mavadia2017} considers a weighted sum of past measurements to predict the $i$-th step ahead measurement outcome, $i \in [0, N_P]$. A gradient descent algorithm learns the weights, $\{\phi_{q' \leq q}\}$ for the previous $q$ past measurements, and a constant offset value for non-zero mean processes, to calculate the $i$-th step ahead prediction. The set of $N_P$ LSF models, collectively, define the set of predicted qubit states under an LSF acting on a measurement record.  For $i=1$, equivalent to the single-step update employed in the Kalman filter, we assert that learned $\{\phi_{q' \leq q}\}$ in LSF effectively implements an AR($q$) process (we validate numerically in Section~\ref{sec:main:Performance}). Under this condition, and for zero-mean $w_n$, the LSF in \cite{mavadia2017} by definition searches for coefficients for the weighted linear sum of past $q$ measurements, as described in in Eq.~(\ref {eqn:main:ARprocess}). 

We use the parameters $\{\phi_{q' \leq q}\}$ learned in the LSF to define $\Phi$ in Eq.~(\ref {eqn:akf_Phi}), therefore reducing the computational complexity of the remaining optimisation from ($(q+2)\to 2$)-dimensional for an AKF of order $q$. Since Kalman noise parameters ($\sigma^2, R$) are subsequently auto-tuned using a Bayes Risk optimisation procedure (see Section~\ref{sec:main:Optimisation}), we optimise over potential remaining model errors and measurement noise.  

In general, LSF performance improves as $q$ increases and a full characterisation of model-selection decisions for LSF are given in \cite{mavadia2017}. Defining an absolute value for the optimal $q$ is somewhat arbitrary as it is defined relative to the extent to which a true $\state$ is oversampled in the measurement routine and the finite size of the data. For all analyses presented here, we fix the ratio $q \Delta t = 0.1 $ [a.u.] and $q / N_T = 0.05$ [a.u.], where the experimental sampling rate is $1/\Delta t$, $N_T$ and $\{\phi_{q' \leq q}\}$ are identical in the AKF and LSF.   In practice this ensures numerical convergence of the LSF during training.

\subsubsection{Liska Kalman Filter with Fixed Basis (LKFFB)}
In LKFFB, we effectively perform a Fourier decomposition of the underlying $\state$ in order to build the dynamic model, $\Phi$, for the Kalman filter.   Here, we project our measurement record on $J^{(B)}$ oscillators with fixed frequency $\omega_{j}\equiv j\omega_0^{(B)}$ with $j$ an integer as $j = 1, \hdots, J^{(B)}$. The temporal resolution of the state tracking procedure is set by the maximum frequency in the selected basis and properties of the spacing between adjacent basis frequencies. The superscript $ ^{(B)}$ indicates Fourier domain information about an algorithmic basis, as opposed to information about the true (unknown) dephasing process.  The LKFFB allows instantaneous amplitude and phase tracking for each basis oscillator, directly enabling forward prediction from the learned dynamics.  The structure of this Kalman filter, referred to as the Liska Kalman Filter (LKF), was developed in \cite{livska2007}; adding a fixed basis in this application yields the Liska Kalman Filter with a Fixed Basis (LKFFB). 

For our application, the true hidden Kalman state, $x$, is encoded as a collection of sub-states, $x^j$, for the $j^{th}$ oscillator. For clarity we remind that the superscript is used as an index rather than a power.  Each sub-state is labeled by a real and imaginary component which we represent in vector notation: 
\begin{align}
	x_n & \equiv \begin{bmatrix} x^{1}_{n} \hdots x^{j}_{n} \hdots x^{J^{(B)}}_{n} \end{bmatrix} \\
	A^j_{n} & \equiv \textrm{Re}(x^{j}_{n}) \\
	B^j_{n} & \equiv \textrm{Im}(x^{j}_{n}) \\
	x^j_n & \equiv \begin{bmatrix} A^j_{n} \\ B^j_{n}  \end{bmatrix}
\end{align} 
The algorithm tracks the real and imaginary parts of the Kalman sub-state simultaneously in order calculate the instantaneous amplitudes ($\norm{x^j_n}$) and phases ($\theta^{j}_{n}$)  for each Fourier component:
\begin{align}
	\norm{x^j_n} & \equiv \sqrt{(A^j_{n})^2 + (B^j_{n})^2} \\
	\theta^{j}_{n} & \equiv \tan{\frac{B^j_{n}}{A^j_{n}}}
\end{align}

The dynamical model for LKFFB is now constructed as a stacked collection of these independent oscillators. The sub-state dynamics match the formalism of a Markovian stochastic process defined on a circle for each basis frequency, $\omega_j$, as in Ref.~\cite{karlin1975first}. We stack $\Theta(j \omega_0^{(B)}\Delta t) $ for all $\omega_j$ along the diagonal to obtain the full dynamical matrix for $\Phi_n$:
\begin{align}
	\Phi_{n} & \equiv \begin{bmatrix} 
		\Theta(\omega_0^{(B)}\Delta t)\hdots 0  \\ 
		\hdots \Theta(j\omega_0^{(B)}\Delta t) \hdots \\
		0 \hdots \Theta(J^{(B)} \omega_0^{(B)}\Delta t)  \end{bmatrix}\\ 
	\Theta(j \omega_0^{(B)}\Delta t) &\equiv \begin{bmatrix} \cos(j \omega_0^{(B)}\Delta t) & -\sin(j \omega_0^{(B)}\Delta t) \\ \sin(j \omega_0^{(B)}\Delta t) & \cos(j \omega_0^{(B)}\Delta t) \\ \end{bmatrix} \label{eqn:ap_approxSP:LKFFB_Phi} 
\end{align}

We obtain a single estimate of the true hidden state by defining the measurement model, $H$, by concatenating $J^{(B)}$ copies of the row vector $[1\;\;0]$ :
\begin{align}
	H & \equiv \begin{bmatrix} 1\;\;0 \hdots 1\;\;0 \hdots 1\;\;0 \end{bmatrix}
\end{align}
Here, the unity values of $H$ pick out and sum the Kalman estimate for the real components of $\state$ while ignoring the imaginary components, namely, we sum $A^{j}_{n}$ for all $J^{(B)}$ basis oscillators.

In \cite{livska2007}, a state-dependent process-noise-shaping matrix is introduced to enable potentially non-stationary instantaneous amplitude tracking in LKKFB for each individual oscillator: 
\begin{align}
	\Gamma_{n-1} &\equiv \Phi_{n-1}\frac{x_{n-1}}{\norm{x_{n-1}}}
\end{align}
For the scope of this manuscript, we retain the form of $\Gamma_{n}$ in our application even if true qubit dynamics are covariance stationary. As such, $\Gamma_{n}$ depends on the state estimates $x$. For this choice of $\Gamma_{n}$, we deviate from classical Kalman filters because recursive equations for $P$ cannot be propagated in the absence of measurement data. Consequently, Kalman gains cannot be pre-computed prior to experimental data collection. Details of gain pre-computation in classical Kalman filtering can be found in standard textbooks (e.g. \cite{grewal2001theory}).

There are two ways to conduct forward prediction for LKFFB and both are numerically equivalent for an appropriate choice of basis: (i) we set the Kalman gain to zero and recursively propagate using $\Phi$; (ii) we define a harmonic sum using the basis frequencies and learned $\{\norm{x^j_n}, \theta^{j}_{n} \}$.  This harmonic sum can be evaluated for all future time to yield forward predictions in a single calculation. The choice of basis for an LKFFB and its implications for optimal predictive performance are discussed in Appendix~\ref{sec:app:subsec:LKFFB}.

\subsubsection{Quantised Kalman Filter (QKF)}

In QKF, we implement a Kalman filter that acts directly on discretised measurement outcomes, $d \in \{0,1\}$. To reiterate the discussion of  Fig.~\ref{fig:main:Predive_control_Fig_overview_17_one}(a), this means that the measurement action in QKF must be non-linear and take as input quantised measurement data. This holds true irrespective of our dynamical model, $\Phi$.  In our application we set the dynamical model to be identical to that employed in the AKF, allowing isolation of the effect of the nonlinear, quantised measurement action.

With unified notation across AKF and QKF, we define a non-linear measurement model $h(x)$ and its Jacobian, $H$ as:
\begin{align}
	z_n &  \equiv h(x_n[0]) \equiv \frac{1}{2}\cos(\state_{n}) \\
	\implies H_n &\equiv \frac{d h(\state_n)}{d\state_n} =  -\frac{1}{2}\sin(\state_{n})
\end{align}
During filtering, $z_n = h(x_n[0])$ is used to compute measurement residuals when updating the true Kalman state, $x_n$, whereas the state variance estimate, $P_n$, is propagated using the Jacobian, $H_n$. Further, the Jacobian  is used to compute the Kalman gain. Hence the filter can quickly destabilise if the linearisation of $h(\cdot)$ by $H_n$ doesn't hold during dynamical propagation, resulting in a rapid build up of errors. 

In this construction, the entity $z_n$ is associated with an abstract `signal': a sequence formed by repeated applications of the likelihood function for a single qubit measurements in Eq.~(\ref {eqn:main:likelihood}).  The true stochastic qubit phase, $\state_n$, is our Kalman hidden state, $x_n$. Subsequently, we extract an estimate of the true bias, $z_n$, as an unnatural association of the Kalman measurement model with Born's rule. The sequence $\{z_n\}$ is not observable, but can only be inferred over a large number of experimental runs. 

To complete the measurement action, we implement a biased coin flip within the QKF filter given $\tilde{y}_n$.  While the qubit provides measurement outcomes which are naturally quantised, we require a theoretical model, $\mathcal{Q}$, to generate quantised measurement outcomes with statistics that are consistent with Born's rule in order to propagate the dynamic Kalman filtering equations appropriately. In order to build this machinery we modify the procedure in \cite{karlsson2005} to quantise $z_n$ using biased coin flips. In our notation, we represent a black-box quantiser, $\mathcal{Q}$, that gives only a $0$ or a $1$ outcome based on $\tilde{y}_n$:
\begin{align}
	d_n &= \mathcal{Q}(\tilde{y}_n)\\
	&=  \mathcal{Q}(h(\state_n) + v_n)
\end{align}
The use of the notation $\tilde{y}_n$ is meant to indicate a correspondence with $y_{n}$ introduced earlier, while the physical meaning differs due to the discretised nature of the QKF.  Therefore, the stochastic changes in $\{ \tilde{y}_n\}$ are represented in the bias of a coin flip, subject to proper normalisation constraints which maintains $|\tilde{y}_n| \leq 0.5$:
\begin{align}
	Pr(d_n| \tilde{y}_n, \state_{n}, \tau) & \equiv \mathcal{B}(n_{\mathcal{B}}=1;p_{\mathcal{B}}= \tilde{y}_n + 0.5 ) \label{eqn:main:qkf:binomial}
\end{align}
QKF uses Eq.~(\ref {eqn:main:qkf:binomial}) to define a biased coin-flip during filtering, where $n_{\mathcal{B}}$ represents a single coin flip, $p_{\mathcal{B}}$ represents the stochastically drifting bias on the coin. Kalman filtering with the coin-flip quantisation defined by Eq.~(\ref {eqn:main:qkf:binomial}) presents a departure from classical amplitude quantisation procedures in \cite{widrow1996, karlsson2005}.

From a computational perspective, we modify the process noise features definition from AKF to QKF. We set $Q \equiv \sigma^2\Gamma \Gamma^T \to \sigma^2 \mathcal{I} \quad \forall n $, $\mathcal{I}$ is $q\times q$ identity matrix, from AKF to QKF. This rationale for this modification is that it smears out the effect of white process noise in a way that stabilizes inversions in the gain calculation in the Kalman filter, but does not correlate any two Kalman states in time (diagonal matrix). In practice, this modification only yields mild improvements over the original AKF process noise features matrix.

The definitions of $\{ \mathcal{Q}, h(x_n), H_n, Q \}$ in this subsection, and dynamics $\{x_n, \Phi\}$ from the AKF now completely specify the QKF algorithm for application to a discrete, single-shot measurement record as depicted in Fig.~\ref{fig:main:Predive_control_Fig_overview_17_one} (a).  

\subsection{Gaussian Process Regression (GPR)}

In GPR, correlations in the measurement record can be learned if one projects data on a distribution of Gaussian processes, $Pr(\state)$ with an appropriate encoding of their covariance relations via a kernel, $\Sigma_\state^{n_1, n_2}$. We return to the linear measurement record and the definition of scalar noisy observations $y_{n}$ corrupted by Gaussian measurement noise, $v_n$, as considered previously for AKF, LSF, and LKFFB.  
Under linear operations, the distribution of measured outcomes, $y_n$, is also a Gaussian. The  mean and variance of $Pr(y)$  depends on the mean $\mu_\state$ and variance $\Sigma_\state$ of the prior $Pr(\state)$, and the mean $\mu_v \equiv 0$ and variance $R$ of the measurement noise: 
\begin{align}
	\state & \sim Pr_\state(\mu_\state,\Sigma_\state ) \\
	y & \sim Pr_y(\mu_\state,\Sigma_\state + R ) 
\end{align}
For covariance stationary $\state$, correlation relationships depend solely on the time lag, $\nu \equiv \Delta t|n_1 - n_2|$ between any two time points  $n_1, n_2 \in [-N_T, N_P]$.  An element of the covariance matrix, $\Sigma_\state^{n_1,n_2}$, corresponds to one value of lag, $\nu$, and the correlation for any given $\nu$  is specified by the covariance function, $R(\nu)$:
\begin{align}
	\Sigma_\state^{n_1,n_2} & \equiv R(\nu) 
\end{align}
Any unknown parameters in the encoding of correlation relations via $R(\nu)$ are learned by solving the optimisation problem outlined in Section~\ref{sec:main:Optimisation}. The optimised GPR model is then applied to datasets corresponding to new realisations of $\state$. Let indices $n \in N_T \equiv [-N_T, 0]$ denote training points, and let a length $N^{\ddagger} $ vector contain arbitrary testing points $n^{\ddagger} \in [-N_T, N_P]$. These testing points in machine learning language encompass both state estimation and prediction points in our notation. We now define the joint distribution $Pr(y,\state^{\ddagger})$, where $\state^{\ddagger}$ represents the true process evaluated by GPR at desired test points: 
\begin{align}
	\begin{bmatrix} \state^{\ddagger} \\y \end{bmatrix} & \sim \mathcal{N} (\begin{bmatrix} \mu_{\state^{\ddagger}} \\ \mu_y
	\end{bmatrix} , \begin{bmatrix}   K(N^{\ddagger},N^{\ddagger})&K(N_T,N^{\ddagger}) \\ K(N^{\ddagger},N_T) & K(N_T,N_T) + R \end{bmatrix} )
\end{align}
The additional `kernel' notation $\Sigma_\state  \equiv K(N_T, N_T)$ is ubitiquous in GPR. Time domain correlations specified by $R(\nu)$ populate each element of a matrix $K(\cdot, \cdot \cdot)$, where the dimensions of the matrix depend on the vector length of each argument. For example, for $K(N_T,N_T)$, the notation defines a square matrix where diagonals correspond to $\nu=0$ and off-diagonal elements correspond to separation of two arbitrary points in time i.e. $\nu \neq 0 $. 

Following \cite{rasmussen2005gaussian}, the moments of the conditional predictive distribution $Pr(\state^{\ddagger}|y)$ can be derived from the joint distribution $Pr(y,\state^{\ddagger})$ via standard Gaussian identities:
\begin{align}
	\mu_{\state^{\ddagger}|y} &= \mu_\state + K(N^{\ddagger},N_T)(K(N_T,N_T) + R )^{-1} (y - \mu_y) \\
	\Sigma_{\state^{\ddagger}|y} &= K(N^{\ddagger},N^{\ddagger}) \nonumber \\
	& - K(N^{\ddagger},N_T)(K(N_T, N_T) + R)^{-1}K(N_T,N^{\ddagger}) 
\end{align}
The prediction procedure outlined above holds true for any choice of kernel, $R(\nu)$. In any GPR implementation, the dataset, $y$, constrains the prior model yielding an a posteriori predictive distribution. The mean values of this predictive distribution, $\mu_{\state^{\ddagger}|y}$, are the state predictions for the qubit under dephasing at test points in $N^{\ddagger}$.

In our work we focus on a `periodic kernel' to encode a covariance function which is theoretically guaranteed to approximate any zero-mean covariance stationary process, $\state$, in the mean square limit, by having the same structure as a covariance function for trigonometric polynomials with infinite harmonic terms \cite{solin2014explicit, karlin1975first}. The sine squared exponential kernel represents an infinite basis of oscillators and is defined as:
\begin{align}
	R(\nu) &\equiv \sigma^2 \exp (- \frac{2\sin^2(\frac{\omega_0^{(B)}\nu}{2})}{l^2}) 
\end{align} 
This kernel is described using just two key hyper-parameters: the frequency-comb spacing for our infinite basis of oscillators, $\omega_0$, and a dimensionless length scale, $l$. We use physical sampling considerations to approximate their initial conditions prior to an optimisation procedure, namely, that the longest correlation length encoded in the data sets the frequency resolution of the comb, and the scale at which changes in $\state$ are resolved is limited physically by the minimum time taken between sequential Ramsey measurements:
\begin{align}
	\frac{\omega_0^{(B)}}{2\pi} & \sim  \frac{1}{\Delta t N} \\
	l & \sim \Delta t
\end{align} 
Because the periodic kernel can be shown to be formally equivalent to the basis of oscillators employed in the LKFFB algorithm in a limiting case (see Appendix~\ref{sec:app:spec_methods} for a discussion using results in \cite{solin2014explicit}), the inclusion of GPR using this kernel permits a comparison of the underlying algorithmic structures for the task of predictive estimation using spectral methods.

For the analysis of covariance stationary time series under a GPR framework, we de-emphasise popular kernel choices such as: a Gaussian kernel (RBF),  a scale mixture of Gaussian kernels (RQ), and Matern kernels (e.g. MAT32) \cite{rasmussen2005gaussian,tobar2015learning}. An arbitrary-scale mixture of zero-mean Gaussian kernels will probe an arbitrary area around zero in the Fourier domain, as schematically depicted in Fig.~\ref{fig:main:Predive_control_Fig_overview_17_two}(a). While such kernels capture the continuity assumption ubitiquous in machine learning, they are structurally inappropriate for probing a process characterized by an arbitrary power spectral density (e.g. ohmic noise). Another common kernel for time-series analysis is a quasi periodic kernel (QPER) defined by a product of an RBF with a periodic kernel \cite{roberts2013gaussian}. This corresponds to a convolution in the Fourier domain giving rise to a comb of Gaussians at the expense of an increase in the number of parameters required for kernel tuning. One can also consider specific types of AR($q$) processes using Matern kernels of order $q+1/2$ but with increased restrictions on the form of coefficients \cite{rasmussen2005gaussian,stein2012interpolation}. A simple consideration of autoregressive approaches suggest that a Matern kernel for $q=1$ (MAT32) can be briefly trialed under GPR, whereas high-$q$ autoregressive processes are naturally and generally treated under a KF framework.  Further discussion of kernel choice appears in Sec.~\ref{sec:main:discussion}.

\section{Algorithm Performance Characterisation \label{sec:main:Performance}}

In the results to follow, our metric for characterising performance of optimally tuned algorithms will be the normalised Bayes prediction risk:
\begin{align}
	\normpr \equiv \frac{L_{BR}(n|I)}{\langle \left(\state_n - \mu_\state \right)^2 \rangle_{\state, \mathcal{D}}}, \quad \mu_\state \equiv 0
\end{align}
A desirable forward prediction horizon corresponds to maximal $n^* \in [0, N_P]$ for which normalised Bayes prediction risk at all time-steps $n \leq n^*$ is less than unity. We compare the difference in maximal forward prediction horizons between algorithms in the context of realistic operating scenarios.  We begin here by introducing the numerical methods employed for generating data-sets on which predictive estimation is performed.

We simulate environmental dephasing through a Fourier-domain procedure described in Appendix~\ref{sec:app:truenoise} \cite{soare2014} in order to simulate an $\state$ which is mean-square ergodic and covariance stationary.  For the results in this manuscript, we choose a flat top spectrum with a sharp high-frequency cutoff for simplicity as this choice of a power spectral density theoretically favors no particular choice of algorithm but violates the Markov property. 

In our simulations we also must mimic a measurement process which samples the underlying ``true'' dephasing process.  The algorithmic parameters $\{N_T, \Delta t\} $ represent a sampling rate and Fourier resolution set by the simulated measurement protocol; we choose regimes where the Nyquist rate, $r \gg 2$. In generating noisy simulated measurement records, we corrupt a noiseless measurement by additive Gaussian white noise. Since $\state$ is Gaussian, the measurement noise level, $N.L.$, is defined as a ratio between the standard deviation of additive Gaussian measurement noise, $\sqrt{R}$ and the maximal spread of random variables in any realisation $\state$. We approximate the maximal spread of $\state$ as three sample standard deviations of one realisation of true $\state$, $N.L. = \sqrt{R}/3\sqrt{\hat{\Sigma}_\state^{n,n}}$. The use of a hat in this notation denotes sample statistics. This computational procedure enables a consistent application of measurement noise for $\state$ from arbitrary, non-Markovian power spectral densities. For the case where binary outcomes are required, we apply a biased coin flip using Eq.~(\ref {eqn:main:qkf:binomial}).

\subsection{Algorithmic Optimisation \label{sec:main:Optimisation}}

All algorithms in this manuscript employ machine learning principles to tune unknown design parameters based on training data-sets. The physical intuition associated with optimising our filters is that we are cycling through a large class of general models for environmental dephasing and seeking the model(s) which best fit the data subject to various constraints. This allows each filter to track stochastic qubit dynamics under arbitrary covariance-stationary, non-Markovian dephasing.  We elected to deploy an optimisation routine with minimal computational complexity to enable nimble deployment of KF and GPR algorithms in realistic laboratory settings, particularly since LSF optimisation is extremely rapid for our application \cite{mavadia2017}. 

Kalman filtering in our setting poses a significant challenge for general optimisers as the lack of theoretical bounds on the values of ($\sigma, R$) result in large, flat regions of the Bayes Risk function. Further, the recursive structure of the Kalman filter means that no analytical gradients are accessible for optimising a choice of cost function and a large computational burden is incurred for any optimisation procedure. We randomly distribute $(\sigma_{k}, R_{k})$ pairs for $k=1, \hdots K $ over ten orders of magnitude in two dimensions in order to sample the optimisation space.

We then generate a sequence of loss values $L(\sigma_k, R_k)$ for each $k$ by considering a small region around $n=0$, where the size of the region is $n_L$ number of time steps we look forward or backwards from $n=0$:
\begin{align}
	L(\sigma_k, R_k) \equiv  \sum_{n=1}^{n_L} L_{BR}(n | I= \{\sigma_k, R_k \}) \label{eqn:main:risk:optriskvalue}.
\end{align}
Here, $L_{BR}(n | I= \{\sigma_k, R_k \})$ is given by Eq.~(\ref {eqn:main:sec:ap_opt_LossBR}) and it is summed over $0\leq n_L\leq |N_{T}|$  ($0\leq n_L\leq |N_{P}|$) backwards (forwards) time-steps for state estimation (prediction). In the notation for $I$ above, we omit Kalman dynamical model design parameters for ease of reading.  Typically $I$ would include, for instance, the set of autoregressive coefficients in AKF and the set of fixed basis frequencies in LKFFB. Values of $n_L$ are chosen such that the sequence $\{L(\sigma_k, R_k) \}$ defines sensible shapes of the total loss function over parameter space and the numerical experiments in this manuscript. A choice of small $n_L$ in state-estimation ensures that data near the prediction horizon are employed - a region where the Kalman filter is most likely to have converged.  Similarly, in state prediction, large $n_L$ will flatten the true prediction loss function as long-term prediction errors dominate smaller loss values occurring during the short term prediction period. In addition, one can weight state estimation and state prediction loss functions differently by choosing different values of $n_L$ for state estimation and prediction, though we set $n_L$ to be the same in both regions. While simple and by no means optimal, our tuning approach is computationally tractable and efficient compared to the application of standard optimisation routines where each loss value calculation requires a recursive filter to act on a long measurement record. Further, our approach ensures tuning procedures are performed off-line such that a tuned algorithm is simple in its recursive structure and performs rapid calculations at each time-step.

An ideal parameter pair ($\sigma^*, R^*$) minimises Bayes risk over $K$ trials for both state estimation and prediction.  We define acceptable low loss regions for state estimation and prediction as being the set which returns loss less than 10\% of the median risk over $K$ trials.  In the event that low risk regions do not exist for both state estimation and prediction for a given parameter pair, we deem the optimisation to have failed as state estimation performance is uncorrelated with forward prediction (for illustration, see panel (h) of Fig.~\ref{fig:main:fig_data_specrecon}).

In GPR the set of parameters $I = \{\sigma, R, \omega_0^{(B)}, l \}$ requires optimisation.  However, in contrast to the KF, no recursion exists and analytic gradients are accessible to simplify the overall optimisation problem. Instead of minimising Bayes state-estimation risk, we follow a popular practice of maximising the Bayesian likelihood. Initial conditions and optimisation constraints are derived from physical arguments as described in Section~\ref{sec:main:OverviewofPredictive Methodologies}.

\subsection{Performance of the KF using linear measurement}
\begin{figure}
	\includegraphics[scale=1.0]{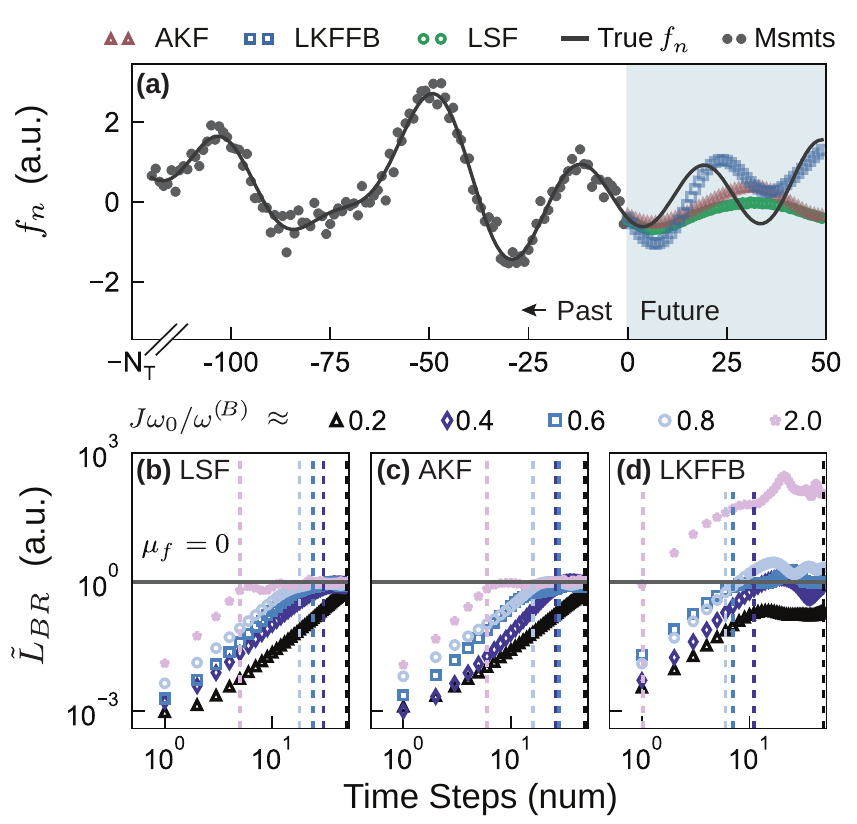}
	\caption{\label{fig:main:fig_data_state_pred} (a) Solid dots depict $y_n$ against time-steps $n$ and data collection ceases at $n=0$. Optimised LSF, AKF and LKFFB yield predictions $n>0$ in the blue region plotted as open, coloured markers. A black solid line shows one realisation of true $\state_n$, drawn from a flat top spectrum with $J$ true Fourier components spaced $\omega_0$ apart and uniformly randomised phases. Other parameters: $\omega_0 / \omega_0^{(B)}  \notin \mathcal{Z}$ (natural numbers), $J = 45000$, $\omega_0 / 2\pi = \frac{8}{9} \times 10^{-3}$ Hz such that $>500$ number of true components fall between adjacent LKFFB oscillators; $ N.L.= 10\%$. (b)-(d) Procedure in (a) is repeated for ensemble $M$ different realisations of $\state$ and noisy datasets to compute $\normpr$ for LSF, AKF, and LKFFB. $\normpr$ v. $n \in [0, N_P]$ is plotted; dark-grey horizontal line marks $\normpr \equiv 1$ for predicting the mean $\mu_\state \equiv 0$. Vertical dashed lines mark the forward prediction horizon, $ n^* $, where $  \normpr \lesssim 0.8 < 1$ for all prediction time steps  $0< n \leq n^*$ in out-performing predicting the noise mean. Marker color (dark indigo to pink) depicts true $\state$ cutoff, $J\omega_0$ varied relative to $\omega^{(B)} \equiv \omega_0^{(B)}J^{(B)} \approx r \omega_{(S)}$, with fixed Nyquist $r\gg2$; $\omega_0 / 2\pi = 0.497$ Hz, $J = 20, 40, 60, 80, 200$; $N.L. = 1\%$.  For all (a)-(d), a trained LKFFB is implemented with $\omega_0^{(B)} / 2\pi = 0.5$ Hz and $J^{(B)} =100$ oscillators; trained AKF / LSF models are $q = 100$; with $N_T = 2000, N_P = 50$ steps, $\Delta t = 0.001$s, $M=50$ runs, $K=75$ optimisation trials.} 
\end{figure}

The general performance of the various KF algorithms discussed above is illustrated in Fig.~\ref{fig:main:fig_data_state_pred} which compares the AKF and LKFFB algorithms using a linear measurement record.  Here the solid black line represents the underlying true $\state$ and solid markers indicate noisy simulated linear measurement data.  Future predictions using the various KF formalisms and the (non-recursive) LSF filter~\cite{mavadia2017} are shown as coloured open markers, based on these data.  The selected single realization of the prediction process demonstrated in (a) is representative of a broad ensemble of simulated data sets and demonstrates the ability of all algorithms to perform future prediction with varying degrees of success.

In general, our objective is to maximise the forward prediction horizon, $n^{*}$, in any algorithmic setting.  In Fig.~\ref{fig:main:fig_data_state_pred}(b)-(d), we explore the key determining factors setting the value of the prediction horizon under the three main Kalman filtering algorithms treated here.  We plot the ensemble-averaged $\normpr$ as a function of forward prediction time when adjusting the ratio of the cutoff frequency in the noise, $J\omega_{0}$, to the sample rate in the measurement routine ($\omega_{(S)}=2\pi/\Delta t$) without physical aliasing such that Nyquist $r \gg2$ and $\omega_{(S)} \approx \omega^{(B)} / r$, where $\omega^{(B)}$ incorporates a (potentially incorrect) bandwidth assumption about dephasing noise for LKFFB. Here again, we have a forward prediction horizon for time-steps $0 < n < n^*$ if $\normpr\lesssim 1$ for all time-steps in this region and an algorithm seeks to maximise $n^*$. In this region, each algorithm predicts future dynamics better than naively predicting the mean behaviour of $\state$ ($\mu_\state \equiv 0$), indicated by a dark-grey horizontal line.

The prediction horizon, indicated approximately by dashed vertical lines, for all algorithms increases as the measurement becomes sufficiently fast to sample the highest frequency dynamics of $\state$.  We confirm numerically that absolute prediction horizons for any algorithm are arbitrary and adjustable through the sample rate, allowing us to restrict our analysis to comparative statements between algorithms for future results.  While differences between protocols appear reasonably small we note that in most cases examined the AKF demonstrates superior performance to the LKFFB subject to the realistic constraint that the true dynamics of $\state$ cannot be perfectly projected onto the basis used in LKFFB (the latter situation corresponding to substantial a priori knowledge of the dynamics of $\state$).  The role of undersampling in the LKFFB becomes pronounced as predictive estimates lead to unstable behavior relative to the naive prediction of $\mu_\state = 0$ in the case $J\omega_{0}/\omega^{(B)}=2$ in Fig.~\ref{fig:main:fig_data_state_pred}(d).  The AKF and LSF share autoregressive coefficients and therefore both algorithms demonstrate comparable $\normpr$ prediction risk in the ensemble average.

\begin{figure}[b]
	\includegraphics[scale=1.]{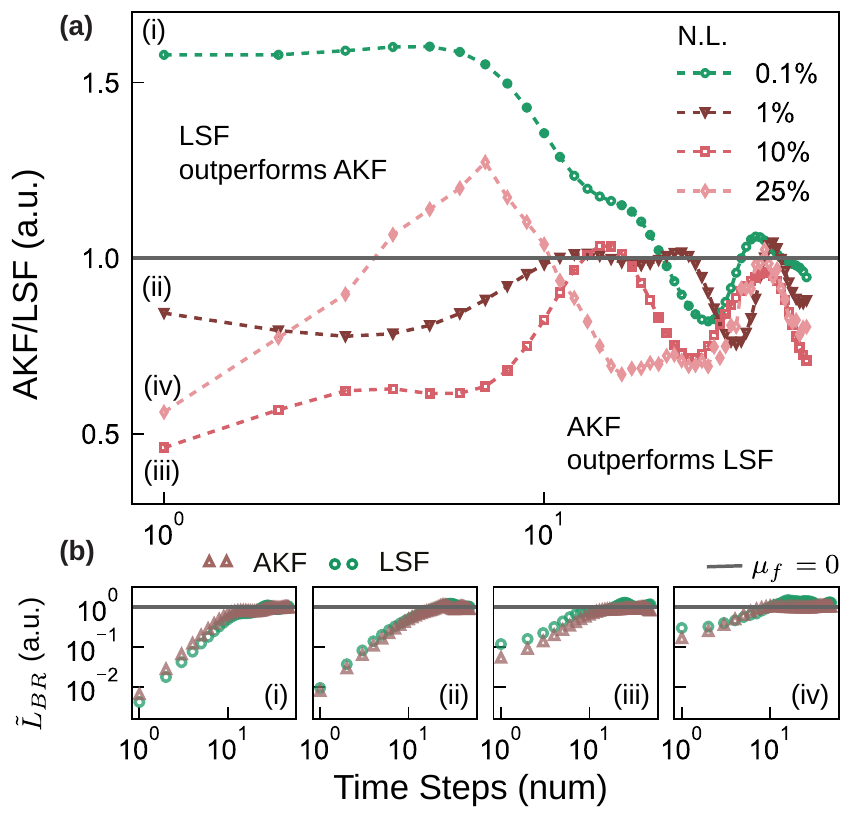}
	\caption{\label{fig:main:fig_data_akfvlsf} Measurement noise filtering in AKF v. LSF. (a) Dashed-lines with markers depict the ratio of $\normpr$ for AKF to LSF against time-steps $n>0$; for cases (i)-(iv) with $N.L. = 0.1, 1.0, 10.0, 25.0 \%$. Green trajectory shows LSF outperforms AKF with ratio $>1$ for $n\leq n^*$; crimson trajectories show AKF outperforms LSF with ratio $<1$ for $n\leq n^*$. (b) $\normpr$ against $n$ is plotted for cases (i)-(iv) confirms a maximal forward prediction horizon marked by $n^*$, exists for all ratios in (a) for both LSF and AKF. In (a) and (b), AKF and LSF share identical $\{ \phi_q \}$. True $\state$ is drawn from a flat top spectrum with $\omega_0 / 2\pi = \frac{8}{9} \times 10^{-3}$ Hz, $J = 45000$, $N_T = 2000, N_P = 100$ steps, $\Delta t = 0.001s, r=20$ such that Fig.~\ref{fig:main:figure_lkffb_path}(c) corresponds to case (ii) in this figure. AKF is optimised with $q=100$, $M=50$ runs, $K=75$ trials.}
\end{figure}

\begin{figure*} [htp]
	\includegraphics[scale=1.0]{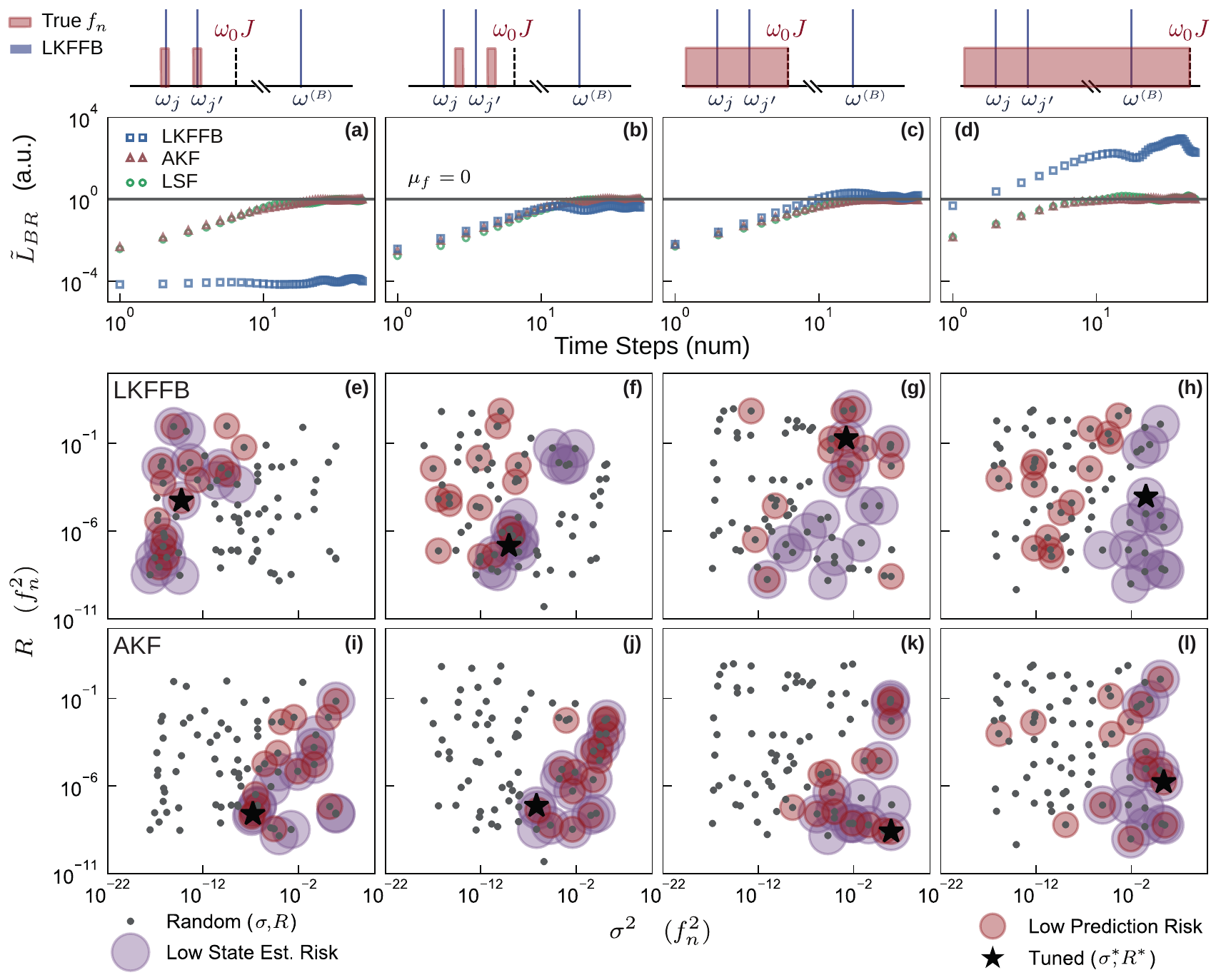}
	\caption{\label{fig:main:figure_lkffb_path} 
		Comparison of KF performance under various imperfect learning scenarios. (a)-(d) True noise properties are varied to introduce pathological learning with respect to fixed algorithmic configuration: $\omega_0 / 2\pi = 0.5, 0.499, \frac{8}{9} \times 10^{-3}, \frac{8}{9} \times 10^{-3}$ Hz and $J = 80, 80, 45000, 80000$ respectively. The relationship between LKFFB basis and true noise spectrum is shown schematically above columns: (a) perfect learning; (b) imperfect projection on LKFFB basis; (c) finite computational Fourier resolution; (d) relaxed basis bandwidth assumption. (a)-(d) $\normpr$  against time-steps $n>0$ is shown for LKFFB, AKF, and LSF. (e)-(l) Optimisation results for LKFFB [top row] and AKF [bottom row] in each of the four regimes in (a)-(d). Grey dots depict $K$ random ($\sigma^2, R$) pairs; where $M$ realisations of $\state, \mathcal{D}$ are used to calculate $\normpr$ for each pair. Purple (crimson) circles represent low loss regions where risk value in Eq.~(\ref {eqn:main:risk:optriskvalue}), for  ($\sigma^2, R$) is $< 10\%$ of the median  risk value during state estimation (prediction) for $-n_L < n <0$ ($n_L > n >0$),  with $n_L=50$. Black star, ($\sigma^*, R^*$), minimises risk values over purple circles during state estimation.  A KF filter is `tuned' if optimal ($\sigma^*, R^*$) lies in the overlap of low loss regions for state estimation [purple] and prediction [crimson]; disjoint regions in (h) show LKFFB tuning failure. KF algorithms set up with $q = 100$ for AKF; $J^{(B)} = 100, \omega_0^{(B)} / 2\pi = 0.5$ Hz for LKFFB; with $N_T = 2000, N_P = 100$ steps, $\Delta t = 0.001$s, $r=20$; $ N.L. = 1\%$.}  
\end{figure*}

A key implied benefit of the use of Kalman filtering vs the LSF with high-order autoregressive dynamics alone is the addition of robustness against measurement noise.  In order to probe this numerically, we perform direct comparisons of filter performance under varying measurement-noise strength for both the AKF and LSF.  Since autoregressive coefficients learned in (noisy) environments are re-cast in Kalman form, we test measurement-noise filtering in Kalman frameworks enabled by the design parameter $R$. In Fig.~\ref{fig:main:fig_data_akfvlsf} (a), we plot $\normpr$ prediction risk for AKF and LSF as a ratio such that a value greater than unity implies LSF outperforms AKF. In cases (i)-(iv), we increase the applied noise level to our data-sets $\{ y_n \}$ representing simulated measurements on $\state$. For applied measurement noise level $N.L. > 1\%$ in (ii)-(iv), we find that $AKF/LSF <1 $ and AKF outperforms LSF for the conditions studied here, with a general trend towards increasing benefits as noise increases until the noise becomes so large (iv) that the benefits fluctuate as a function of $n$. Calculations of the ensemble-averaged $\normpr$ in Fig.~\ref{fig:main:fig_data_akfvlsf} (b) demonstrate that all ratios reported in (a) correspond to a useful forward prediction horizon.

In machine learning or optimal control settings, the robustness of the learning procedure to small changes in the underlying system is an essential characteristic of the algorithm.  In our case, we have already seen that the quality of projection of the true dynamics of $\state$ onto the LKFFB basis can have a significant impact on the quality of learning and predictive estimation.  We explore this initial finding in more detail.  

In Fig.~\ref{fig:main:figure_lkffb_path}, we simulate various learning conditions including (a) perfect learning in LKFFB; (b) imperfect projection relative to the LKFFB basis; (c) imperfect projection combined with finite algorithm resolution; and (d) imperfect learning and undersampling relative to true noise bandwidth. The ordering of figure presentation highlights the degree of impact of the introduced pathologies on LKFFB.  By contrast we find reasonable model robustness in AKF/LSF at the expense of performance in the somewhat unrealistic perfect learning case.  

We expose the underlying optimisation results for choosing an optimal $(\sigma^*, R^*)$ for LKFFB in Fig.~\ref{fig:main:figure_lkffb_path} (e)-(h) and for AKF in Fig.~\ref{fig:main:figure_lkffb_path} (i)-(l). Individual sample points are highlighted as solid dots while low-loss pairs in this 2D space are highlighted for giving low state-estimation [purple] or prediction [crimson] risk via shaded circles.  As the model pathologies indicated above increase, these data demonstrate a divergence between regions of the optimisation space which permit low-loss state estimation and forward prediction for LKFFB.  In contrast, overlap of low loss Bayes Risk regions do not change for AKF across Fig.~\ref{fig:main:figure_lkffb_path} (i)-(l).

\begin{figure}[tp]
	\includegraphics[scale=1.0]{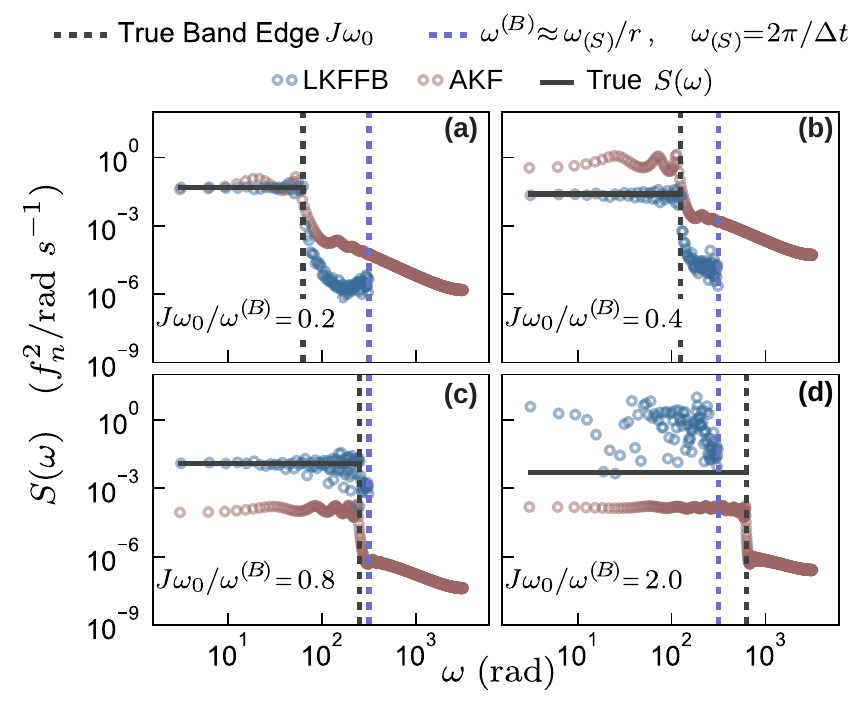}
	\caption{\label{fig:main:fig_data_specrecon} (a)-(d) Blue (red) open markers plot LKFFB (AKF) spectrum estimates; true spectrum (flat top) of $\state$ plotted in black solid line. Dashed black vertical line marks true noise cutoff, $J\omega_0$, and this is varied relative to a measurement sampling rate, $\omega_{(S)}$, and $\omega^{(B)}\equiv \omega_0^{(B)} J^{(B)} \approx \omega_{(S)}/r $ in LKFFB; such that $\omega_0 / 2\pi = 0.497$ Hz, $J = 20, 40, 80, 200$. For LKFFB, blue open markers are $\propto ||\hat{x}^j_n||^2 $ in a single run with $\omega_0^{(B)} / 2\pi = 0.5$ Hz for $j \in J^{(B)} = 100$ oscillators; dashed blue vertical line marks edge of LKFFB basis. For AKF,  red markers are $\hat{S}(\omega)$ computed using learned $\{\phi_{q' \leq q}\}$ and optimised $\sigma^*$, with order $q = 100$.  In all plots, the zeroth Fourier component is omitted on the log scale; and $N_T = 2000, N_P = 50$ steps, $\Delta t = 0.001s, r=20$, with $M=50$ runs, $K=75$ trials; $N.L. = 1\%$.} 
\end{figure}

Kalman filtering algorithms employed here combine recursive state estimation with the establishment of a dynamical model in the Fourier domain.  Therefore, one way to explore algorithmic performance is to look directly at the efficacy of spectral estimation relative to the true (here numerically engineered) hidden dynamics of $\state$.  For both the LKFFB and AKF we plot the extracted power spectral density, $S(\omega)$, as a function of angular frequency $\omega$, for different measurement sampling conditions in Fig.~\ref{fig:main:fig_data_specrecon} against the true spectrum used to define $\state$.  These simulated experimental conditions match those introduced in Fig.~\ref{fig:main:fig_data_state_pred} (b). 

In the case of LKFFB, we plot the learned instantaneous amplitudes from a single run [blue markers] and for AKF we extract optimised algorithm parameters as described above [red markers]. Under the assertion that the LSF implements an AR($q$) process, the set of trained parameters, $\{  \{\phi_{q' \leq q}\}, \sigma^2\}$ from AKF allows us to derive experimentally measurable quantities, including the power spectral density of the dephasing process: $S(\omega) = \sigma^2 \left(2 \pi |1 - \sum_{q'=1}^q \phi_{q'} e^{-i\omega q'}|^2)\right)^{-1} $ \cite{brockwell1996introduction}.  

The critical feature in these data-sets is the existence of a flat-top spectrum possessing a sharp high frequency cutoff.  Both classes of Kalman filtering algorithm successfully identify this structure and locate this high-frequency cutoff.  In general, however, the LKFFB provides superior spectral estimation relative to the AKF, and enables better estimation of the signal strength in the Fourier domain even in the presence of imperfect projection of $\state$ onto the basis used in LKFFB.  The only case in which the LKFFB fails is in Fig.~\ref{fig:main:fig_data_specrecon}(d), where the LKFFB basis is ill-specified relative to the true noise bandwidth. The observed behavior is somewhat surprising given the generally superior performance of the AKF in predictive estimation, but does highlight the practical difference between Fourier-domain spectral estimation and time-domain prediction.

\subsection{Performance of the quantised Kalman filter}
The discrete nature of projective measurement outcomes in quantum systems poses a potential challenge for Kalman filters in the event that measurement pre-processing as in Fig.~\ref{fig:main:Predive_control_Fig_overview_17_one}(b) is not performed.  We test filter performance for predictive estimation when only binary measurement outcomes are available via the QKF.  To re-iterate, QKF estimates and tracks hidden information, $\state_n$, using the Kalman true state $x_n$.  In our construction the associated probability for a projective qubit measurement outcome, $\propto z_n$ is not inferred or measured directly but given deterministically by Born's rule encoded in the non-linear measurement model, $z_n = h(f_n)$. The measurement action is completed by performing a biased coin flip, where $z_n$ determines the bias of the coin.  

\begin{figure}[h!]
	\includegraphics[scale=1.]{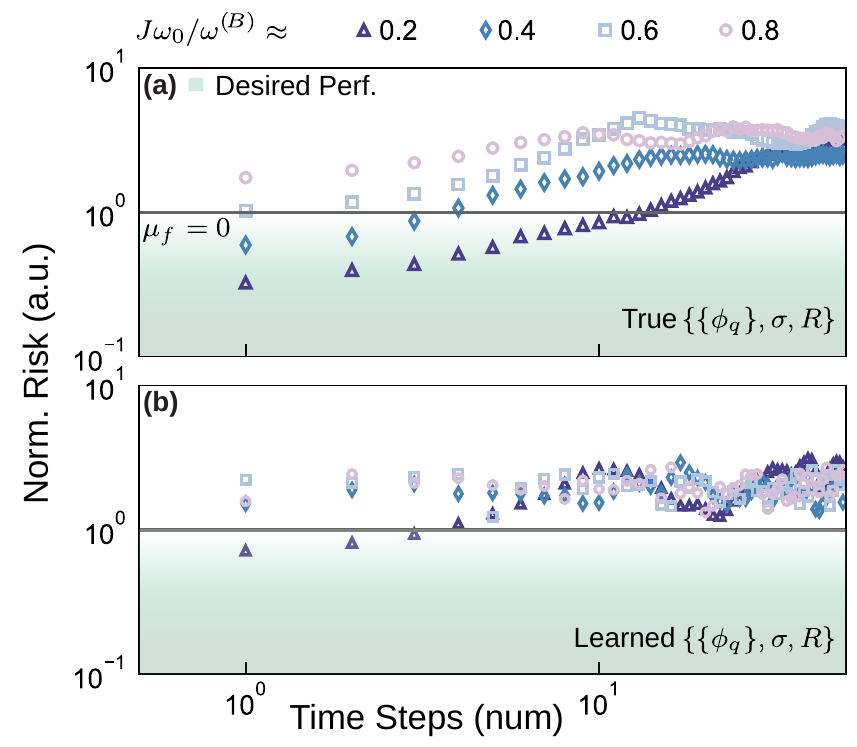}
	\caption{\label{fig:main:fig_data_qkf2}Norm. Risk against $n>0$ plotted for QKF in open markers; dark-grey line at $\mu_\state \equiv 0$ depicts performance under predicting the noise mean. QKF outperforms predicting the mean if open markers lie in green regions. Marker colour (dark indigo to pink) depicts true noise cutoff varied $J \omega_0 / \omega_{(B)} = 0.2, 0.4, 0.6, 0.8$ for $\state$ defined identically in  Fig.~\ref{fig:main:fig_data_specrecon} with $\omega_0/ 2\pi = 0.497 $ Hz, $J = 20, 40, 60, 80$;  $N.L. = 1 \%$. (a) We obtain $\{\phi_{q' \leq q}\}, q=100$ coefficients from AKF/LSF acting on a linear measurement record generated from true $\state$. A new truth, $\state'$, is generated from an AR($q$) process using $\{\phi_{q'\leq q}\}, q=100$ as true coefficients and by defining a known, true $\sigma$. Quantised measurements from $f'$ are obtained; data is corrupted by measurement noise of a true, known strength $R$. (b) We use $\{\phi_{q' \leq q} \}, q=100$ coefficients from (a) but we generate quantised measurements from the original, true $\state$. QKF noise design parameters are optimised for ($\sigma_{AKF}^* \leq \sigma_{QKF}$, $R_{AKF}^* \leq R_{QKF}$) with $M=50$ runs, $K=75$ trials. For (a)-(b), $N_T = 2000, N_P = 50$ steps, $\Delta t = 0.001$s, $r\gg 2$.}
\end{figure}

For QKF, the normalised ensemble-averaged prediction risk, $ \langle (z_n - \hat{z}_n)^2 \rangle_{f, \mathcal{D}} / \langle (z_n - \mu_z)^2 \rangle_{f, \mathcal{D}}$, is calculated with respect to $z$ as the relevant quantity parameterising qubit-state evolution, instead of the stochastic underlying $\state$. This quantity is labeled as Norm. Risk in Fig.~\ref{fig:main:fig_data_qkf2} and we test if $ \langle (z_n - \hat{z}_n)^2 \rangle_{f, \mathcal{D}} / \langle (z_n - \mu_z)^2 \rangle_{f, \mathcal{D}} < 1$ for $0< n < n^*$ can be achieved for numerical experiments considered previously in the linear regime. In particular, we generate true $\state$ defined in numerical experiments in Fig.~\ref{fig:main:fig_data_state_pred}(b) (and Fig.~\ref{fig:main:fig_data_specrecon}) for $q=100$ and varying sample rates.

We isolate the role of the measurement action by first inputting into the QKF a true dynamical model rather than a dynamical model learned as in the standard AKF.  To specify true dynamics, we begin with a set of $\{ \phi_{q'\leq q}\}$ and exactly derive a new $f'$.  As a result the full set of parameters relevant to the filter, $\{\{\phi_{q' \leq q} \}, \sigma, R\}$, are perfectly defined and known, and the filter simply acts on single shot qubit measurements.  These simulations reveal that subject to generic measurement oversampling conditions introduced above the QKF is able to successfully enable predictive estimation.  As in the linear case, the absolute forward prediction horizon is arbitrary relative to $\omega_0 J / \omega^{(B)}$ and implicitly, an optimisation over the choice of $q$ for a finite data size, $N_T$,  in our application. 

Our simulations reveal that the QKF is considerably more sensitive to measurement noise, model errors, and the degree of undersampling than the linear model as shown in Fig.~\ref{fig:main:fig_data_qkf2} (b). Here the QKF incorporates a learned dynamical model from AKF in the linear regime and we tune $(\sigma, R)$ for use in the QKF.  In particular, we explore $\sigma \geq \sigma_{AKF}^*$ to incorporate model errors as $\{\phi_{q' \leq q}\}$ were learned in the linear regime.  We also incorporate increased measurement noise via $R \geq R_{AKF}^*$ as QKF receives raw data that has not been pre-processed or low-pass filtered. The underlying optimisation problems are well behaved for all cases in Fig.~\ref{fig:main:fig_data_qkf2}(b) [not shown].  As the sampling rate is reduced, the QKF forward prediction horizon collapse rapidly i.e $\langle (z_n - \hat{z}_n)^2 \rangle_{f, \mathcal{D}} / \langle (z_n - \mu_z)^2 \rangle_{f, \mathcal{D}} > 1 $ prediction risk for all $n>0$.

\subsection{Failure of GPR in predictive estimation} 
\begin{figure}
	\includegraphics[scale=1.]{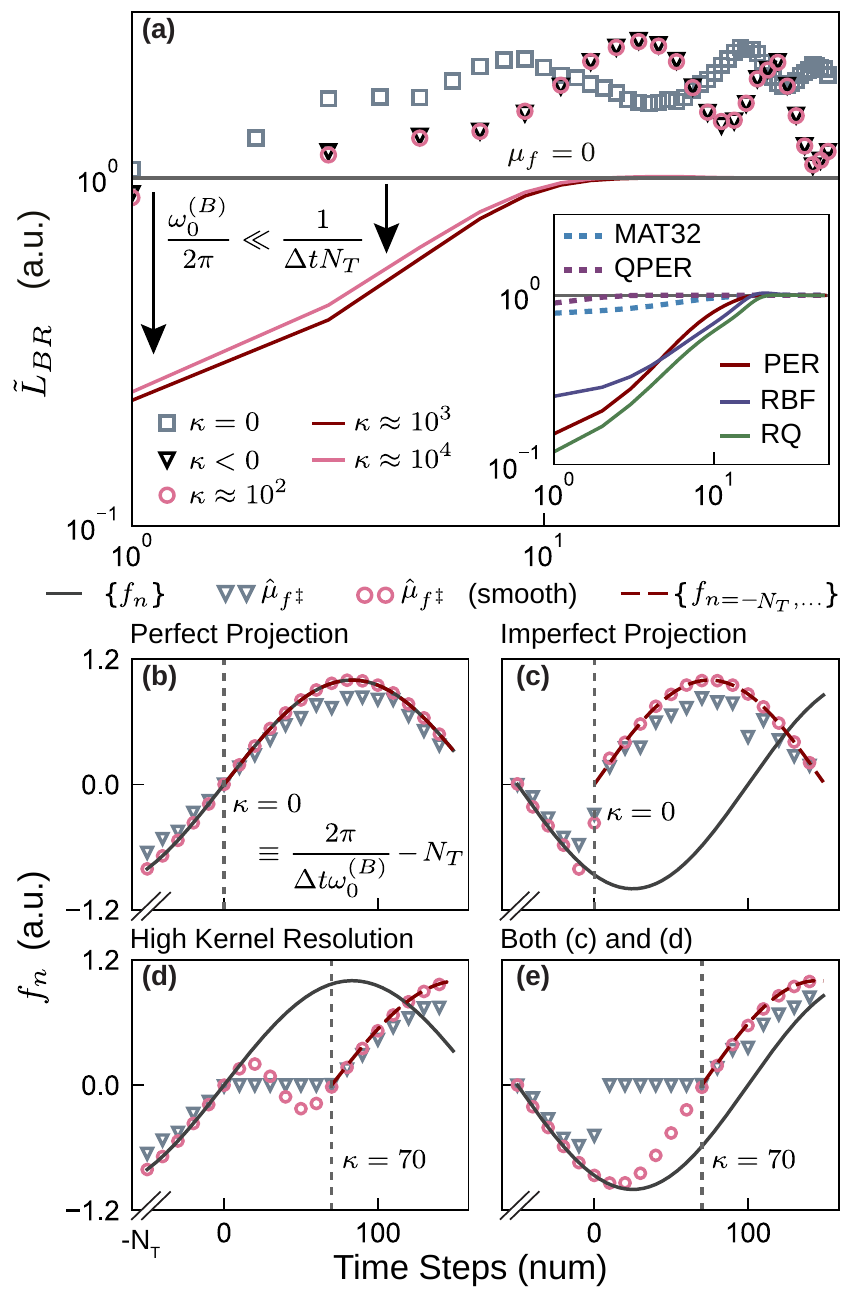}. 
	\caption{\label{fig:main:fig_data_gpr} (a) $\normpr$ v. $n^\ddagger$ (in units of number of time steps) are plotted for GPR with a periodic kernel. Dark-grey horizontal line at unity for $\mu_\state \equiv 0$ marks $\normpr$ under predicting the mean; GPR outperforms predicting the mean if data falls below this line. Grey-black markers correspond to optimisation within physical bounds for $\kappa \leq 0 $ (kernel resolution at or above Fourier resolution); crimson markers and lines depict optimisation within unphysical regimes, $\kappa >0$; with solid lines in high $\kappa \gg 0$ regime. Remaining  $\{R, \sigma, l\}$ optimised for non-negative values. Inset (a) $\normpr$ v. $n^\ddagger$ of periodic kernel (PER) with $\kappa \approx 10^3$ is plotted against results from naively trained Gaussian kernels (RBF, RQ); a Matern kernel (MAT32) and a quasi-periodic kernel (QPER). (b)-(d) True state $\state_n$ v. $n$ [black solid line] and GPR predictions $\hat{\mu}_{\state^\ddagger}$ v. $n^\ddagger$ [open markers]  plotted for periodic kernel for tracking a sinusoid with frequency, $\omega_0$; noisy data record [not shown] ceases at $n=0$. We fix $\kappa = 0, 70$; triangles plot predictions for manually tuned $\{R, \sigma, l\}$; circles plot predictions for optimised $\{R, \sigma, l\}$. Vertical dashed lines mark $n=\kappa$, where we overlay true $\state$ at the beginning of the data record as a red dashed line. (b) Perfection projection is possible $\omega_0 / \omega_0^{(B)} \in \mathcal{Z}$ (natural numbers), $\omega_0/2\pi = 3$ Hz. (c) Imperfect projection, with $\omega_0 / \omega_0^{(B)} \notin \mathcal{Z}$, $\omega_0 / 2 \pi = 3 \frac{1}{3}$ Hz, $\kappa=0$. (d) Moderately raise $\kappa > 0 $, such that $\omega_0 / \omega_0^{(B)} \gg 0 \notin \mathcal{Z}$ for original $ \omega_0 / 2 \pi = 3$ Hz. (e) Test (c) and (d) for $\kappa > 0$, $ \omega_0 / \omega_0^{(B)} \notin \mathcal{Z}$, $\omega_0 / 2 \pi = 3 \frac{1}{3}$ Hz. For (b)-(e), $N_T = 2000, N_P = 150$ steps, $\Delta t = 0.001$s; $N.L.= 1\%$.}    	
\end{figure}

Under a GPR framework, we test whether predictive performance can be improved by considering the entire measurement record (at once) and projecting this record on an infinite basis of oscillators summarised by a periodic kernel. We investigate several different types of GPR models for $M=50$ realisations of $\state$ in the top panel of Fig.~\ref{fig:main:fig_data_gpr}. For the results shown, we use a popular choice of a maximum-likelihood optimisation procedure implemented via L-BFGS in GPy \cite{gpy2014}.

We find that the underlying optimisation procedure for training on our measurement records remains difficult despite having access to an analytical calculation for the cost function. For all results in Fig.~\ref{fig:main:fig_data_gpr}(a) and (b), we use significant manual tuning prior to deploying the automated procedures in GPy. Hence, we focus on using numerical results under GPR to illuminate structural implications of the choice of kernels in our application, rather than making comparative statements about kernel performance.

The results we have assembled demonstrate that the implementation of GPR with a periodic kernel critically depends on the frequency basis comb spacing, $\omega_0^{(B)}$, or equivalently, a deterministic quantity, $\kappa$:
\begin{align}
	\kappa & \equiv \frac{2\pi}{\Delta t \omega_0^{(B)}} - N_T 
\end{align}
The term $ 2\pi /  \Delta t \omega_0^{(B)}$ is the theoretical number of measurements that, in principle, would be required to \emph{physically} achieve the Fourier resolution set by the kernel hyper-parameter, $\omega_0^{(B)}$, and the fundamentally discrete nature of a sequential Ramsey measurement record, expressed by $\Delta t$. Hence, if $\kappa = 0$, the physical Fourier resolution determined by the data set matches the comb spacing in the periodic kernel. For $\kappa > 0$, the comb spacing in the periodic kernel is less than the Fourier spacing defined by the experimental data collection protocol, with total measurements $N_T$. 

In Fig.~\ref{fig:main:fig_data_gpr}(a), we see that GPR predictive performance for the periodic kernel improves as the Kernel's comb spacing is reduced. For each value of $\kappa$ we plot $\normpr$ against time-steps forward, $n^\ddagger$, where the $\ddagger$ corresponds to the evaluation of a predictive GPR distribution on arbitrarily chosen test points, $n^\ddagger = -N_T, \hdots, -1, 0, 1 \hdots, N_P$. Here, the optimiser is constrained to a region in $2\pi/ \omega_0^{(B)}$ parameter space that corresponds to the order of magnitude for $\kappa$. Grey markers correspond to $\kappa \leq 0$, where the algorithm operates above (or at) the Fourier resolution.  In this physically motivated parameter regime, prediction fully fails.  It is not until we set $\kappa\sim10^{3}$ -- a nominally unphysical operating regime where the algorithm's frequency-comb spacing is smaller than the Fourier resolution -- that prediction succeeds [red traces]. This latter case is physically difficult to interpret given that in this regime we find the best ensemble-averaged predictive performance only by providing unphysical freedom to the algorithm.  We note that the optimised length scale for the periodic kernel remains on of order $ \Delta t \sim 10 \Delta t$, such that for all red trajectories in panel (a), we are operating in a high $2\pi/ \omega_0^{(B)}$, low $l$ limit. 

We contextualise the predictive performance of the GPR periodic kernel (PER) [red solid line] in the high-$\kappa$, low-$l$ limit by comparing against predictions derived using other standard kernels [dotted lines] in the inset to ~Fig.~\ref{fig:main:fig_data_gpr}(a). In such circumstances the predictive performance of the periodic kernel predictive is on par with an application of a Gaussian kernel (RBF) and a scale mixture of zero mean Gaussians with different decay lengths (RQ).  A Matern kernel (MAT32) and a quasi periodic kernel (QPER) yield lower-than-anticipated performance. Further discussion of the choice of kernel appears in Sec.~\ref{sec:main:discussion}. For each individual time-trace contributing to the ensemble averages appearing here, we observe that all kernels (PER, RBF, RQ, MAT32, QPER) yield good state estimation and the state estimate at $n^\ddagger=-1$ agrees well with the truth. For GPR with a PER, RBF, and RQ kernels, the state estimate at $n^\ddagger=-1$ smoothly decays to the mean value (zero) for $n^\ddagger \geq 0$ and this effect yields a favourable normalised Bayes prediction risk immediately after $n^\ddagger>0$ depicted by the solid lines in inset of Fig.~\ref{fig:main:fig_data_gpr}(a).

In order to illustrate the operating mechanism for the periodic kernel, we dramatically simplify the model used for $\state$  in  Fig.~\ref{fig:main:fig_data_gpr} (a) and replace it with a single-frequency sine curve.  Fig.~\ref{fig:main:fig_data_gpr} (b)-(e) demonstrates the prediction routine for GPR using a periodic kernel on a simplified version of $\state$, and as before, prediction is always conducted from time-step zero. For this simple example, the periodic kernel learns Fourier information in the measurement record enabling interpolation using test-points $n^{\ddagger} \in [-N_T, 0]$ for all cases (b)-(e) in Fig.~\ref{fig:main:fig_data_gpr}, and atypical features are seen only for test-points in the prediction region [blue shaded region]. We consider predictions from a manually tuned model [triangles] and an optimised GPR model where remaining free $\{\sigma, R, l \}$ parameters are tuned using GPy [circles]. 

An examination of different cases for imperfect learning reveal that this discontinuity exhibits deterministic behavior linked to the underlying structure of the algorithm, namely, to the value of $\kappa$. In our numerical experiments, we find that in all cases of imperfect learning under GPR with a periodic kernel, a discontinuity in the prediction sequence arises at  $n^\ddagger = \kappa$. This is marked by the vertical dashed lines in all panels of Fig.~\ref{fig:main:fig_data_gpr}(b)-(e).  However, another feature appears which we identify as being linked to oversampling of the underlying process determining $\state$.  In such cases, the algorithm simply predicts zero out to $n^\ddagger=\kappa$ before discontinuously predicting future evolution which does not appear similar to the true value of $\state$.  By contrast an optimised model gives smoothly varying predictions, which still adhere to the underlying behaviour set by $\kappa$ for $n^\ddagger>0$. 

In Fig.~\ref{fig:main:fig_data_gpr}(b)-(e), we also plot the value of $\state$ as given from $n=-N_{T}$, the start of the data set, on top of the prediction from $n^\ddagger=\kappa$.  Here we see that the prediction provided by GPR matches the earliest stages of the underlying data set well.  Through various numeric experiments we find that the action of GPR in such parameter regimes (moderately positive $\kappa >0$) appears to be to simply repeat the learned values of $\state$ from $n=-N_{T}$ beginning at $n^\ddagger=\kappa$.  Accordingly these predictions rarely describe the underlying forward dynamics of $\state$ well. 

As we enter the high $\kappa$ regime,  $\kappa  \gg 0$, the features in Fig.~\ref{fig:main:fig_data_gpr}(b)-(e) disappear, and GPR predictions begin to track the (slow moving) `truth' for $n^\ddagger \gg 0$. Analogously to inset (a), we see the performance of PER approach that of standard Gaussian kernels in this simplified case.

\section{Discussion} \label{sec:main:discussion}
The numeric simulations we have performed probe a wide variety of operating conditions in order to explore the algorithmic pathologies of leading forecasting techniques drawn from engineering, econometrics, and machine learning communities.  Our central finding is that overall the autoregressive Kalman filter provides an effective path to perform both state estimation and forward prediction for non-Markovian qubit dynamics. Recasting dynamics into an AKF filter, importantly, provides model robustness against details of the underlying dynamics as well as filtering of noise that allows it to outperform the simpler LSF in \cite{mavadia2017}.  Measurement noise filtering is enabled in the Kalman framework through the optimisation procedure for $R$ and has a regularising (smoothing) effect. Additionally optimisation of the imperfectly learned dynamical model is provided through the tuning of $\sigma$. The joint optimisation procedure over $(\sigma, R)$ ensures that the relative strength of noise parameters is also optimised.  

AKF has also been demonstrated to work well with discretised projective measurement models via what we refer to as the QKF.  In QKF, we employ single-shot, discretised qubit data while enabling model-robust qubit state tracking and increased measurement noise filtering via the underlying AKF algorithm.  However we find that the QKF is vulnerable to the buildup of errors for arbitrary applications and we provide three explanatory remarks from a theoretical perspective. First, the Kalman gains are recursively calculated using a set of \text{linear} equations of motion which incorporate the Jacobian $H_n$ of $h(x_n)$ at each $n$. All non-linear Kalman filters perform well if errors during filtering remain small such that the linearisation assumption holds at all time-steps. Second, measurements are quantised and hence residuals must be $\{-1, 0, 1 \}$ rather than continuously represented floating-point numbers.  In our case, the Kalman update to $x_n$ at $n$, mediated by the Kalman gain cannot benefit from a gradual reduction in residuals.  A third effect incorporates consequences of both quantised residuals and a non-linear measurement action. In linear Kalman filtering, Kalman gains can be pre-calculated in advance of the acquisition of any measurement data: the recursion of Kalman state-variances $P_{n}$, can be decoupled from the recursion of Kalman state-means, $x_{n}$ \cite{grewal2001theory}.  In our application, quantised residuals affect the Kalman update of $x_{n}$, and further, they affect the recursion for the Kalman gain via the state dependent Jacobian, $H_n$. 

In this context, we demonstrate numerically that the QKF achieves a desirable forward prediction horizon when the build of errors during filtering is minimised, for example, by specifying Kalman state dynamics and noise strengths perfectly, and/or by severely oversampling relative to the true dynamics of $\state$.   At present, we simply interpret our results on the QKF as demonstration that one may in principle track stochastic qubit dynamics using single shot measurements under a Kalman framework.  The QKF also has the benefit, as constructed, of reverting to the AKF if suitable pre-processing of data is performed prior to execution of the iterative state-estimation algorithm.  In common laboratory settings the measurement protocol may be effectively linearised through simple averaging of multiple single-shot measurements, application of Bayesian estimation protocols, or other pre-processing as identified above.  So long as the pre-processing takes place on timescales fast relative to the underlying qubit dynamics, the measurement linearization has no impact other than to change the effective sample rate of the measurements.  Thus it is our view that full implementation of the QKF is not essential if improved optimization routines are not accessible.

It is possible that QKF forward prediction horizons in realistic learning environments can be improved by solving the full $q+2$ optimisation problem for $\{\{ \phi_{q' \leq q}\}, \sigma, R\}$, rather than employing the approach taken in this manuscript. However, this poses its own challenges given the observations we make about the optimisation landscape even for the 2D optimisation problem faced in the AKF.  More sophisticated, data-driven model selection schemes are described for both KF and kernel learning machines (such as GPR) in literature (e.g. \cite{arlot2009data, vu2015understanding}). Beyond standard local-gradient and simplex optimisers, we consider coordinate ascent \cite{abbeel2005} and particle swarm optimisation techniques \cite{robertson2017particle} as promising, nascent candidates and their application remains an open research question. One may also consider switching from a high order AR($q$) to an ARMA model with a smaller number of optimisation parameters. Typically, this is accomplished by incorporating either greater a priori information about the underlying dynamic process in the design of the ARMA model and/or using model-less particle-based / unscented filtering techniques to overcome non-linearities in an ARMA representation (e.g. \cite{dong2009unscented}). The latter set of techniques are well adapted for non-linear models but are likely to require a modification to allow for non-Markovian dynamics (e.g. by designing an appropriate transition probability for otherwise Markov re-sampling procedures); in contrast, a typical recursive ARMA formulation for our application may track temporal correlations but be ill-equipped for non-linear, coin-flip measurements. One expects that a straightforward application of such procedures will be complicated.

Our general results on the use of autoregressive models for building Kalman dynamical models stand in contrast to Fourier-domain approaches in LKFFB and GPR using a periodic kernel;  both show significant performance degradation in cases when learning of state dynamics was imperfect.  In investigating the loss of performance for LKFFB, we find that the efficacy of this approach depends on a careful choice of a \textit{probe} (i.e. a fixed computational basis) for the dynamics of $\state$ capturing the effect of dephasing noise on the qubit.  In the imperfect learning regime of Fig.~\ref{fig:main:fig_data_state_pred} and identically, Fig.~\ref{fig:main:fig_data_specrecon}, LKFFB reconstructs Fourier domain information to a high fidelity across a range of sampling regimes but is outperformed by AKF in the time domain (Fig.~\ref{fig:main:fig_data_state_pred}). Since LKFFB tracks instantaneous amplitude and phase information explicitly for each basis frequency, the loss of LKFFB time-domain predictive performance must accrue from difficulty in tracking instantaneous phase, rather than amplitude, information. 

While difficulty of instantaneous phase estimation is likely to disadvantage the time-domain predictive performance of LKFFB, our results show that a Fourier-domain approach yields high fidelity reconstructions of power spectral density describing $\state$. These reconstructions appear robust against imperfect projection on the LKFFB oscillator basis even as oversampling is reduced. This suggests that an application of LKFFB outside of predictive estimation could be tested against standard spectral estimation techniques in future work.

The challenge in adapting GPR for the task of time-domain predictive estimation has proved more striking.  In our numerical simulations, under conditions comparable to those tested in the AKF, the values of normalised Bayes prediction risk for all GPR models are at least an order of magnitude greater than the comparable performance of the AKF or LKFFB (refer to panel Fig.~\ref{fig:main:fig_data_akfvlsf}(b- ii), equivalently, Fig.~\ref{fig:main:figure_lkffb_path}(c)). This difference is somewhat surprising because in the limit that $\Gamma_n$ is set to the identity in LKFFB and an infinite basis of oscillators in the periodic kernel is truncated at the finite value, $J^{(B)}$, both LKFFB and the GPR-PER are formally equivalent to classical Kalman filtering for a collection of $J^{(B)}$ independent state-space resonators \cite{solin2014explicit}. In this limit, the true $\state$ is described by theoretically identical covariance functions in both KF and GPR frameworks. While we do not operate in this regime, one would expect predictive capabilities of these two algorithms to be comparable. 

In contrast to our observations for the various flavors of KF tested here, we observe that GPR predictions with a periodic kernel are useful for filtering/retrodiction but appear to have limited meaning for forward predictions for time-steps $n= n^\ddagger >0$.  In our application, predictive performance of GPR with a periodic kernel for $\kappa=0$ is shown to yield poor predictive performance over the ensemble average (Fig.~\ref{fig:main:fig_data_gpr}(a)). For the unexpected regime of $\kappa \gg 0$ and relatively small fixed $l$, predictive performance improves and the periodic kernel performs similarly to RBF and RQ. In this a high $\kappa$ and a low $l$ regime, the $\sin$ term of the periodic kernel is slowly moving ($\sin(x) \approx x$) and hence the argument of the exponential in the periodic kernel approximates a Gaussian, reducing to an RBF kernel. Our numerical investigations show that an optimised RQ kernel consistently chooses parameter regimes where an RQ also converges to an RBF.  For the operating regimes pertinent to our application, it appears that the choice of the periodic, RBF, and RQ kernels will produce theoretically equivalent results for forward predictions of the qubit state. In our analysis, these `forward predictions' simply arise from a smoothed decay of state estimates starting from test-point $n^\ddagger=-1$ to the noise mean for test-points $n^\ddagger>0$; and are difficult to interpret compared to their Kalman counterparts.

Our numerical characterisation of the periodic kernel for a simple, noiseless $\state$ demonstrates that this kernel learns Fourier domain amplitude information in a way that is better suited for pattern fitting than forward prediction. The predictive time domain sequence of state estimates is repetitive at $ n=n^\ddagger= \kappa$, and can be interpreted as successful qubit-state predictions only when $\state$ is perfectly learned (no discontinuities appear). When learning is imperfect, however, GPR with a periodic kernel is able to learn Fourier amplitudes to provide good retrodictive state estimates for $n^\ddagger<0$, but forward predictions for $n^\ddagger>0$ typically fail.  Unlike LKFFB, we believe the periodic kernel does not permit actively extracting and updating phase information for each individual basis oscillators at $n^\ddagger= \kappa$.  Since phase information can be recast as amplitude information for any fixed-frequency oscillator, one would naively expect that forward predictions can be improved by increasing $\kappa$ moderately, such that the higher order terms in a series expansion of the $\sin$ term are non trivial and $\sin(x)\approx x$ cannot apply. However, any positive value of $\kappa$ means that we are probing dynamics at frequencies lower than appearing in the data-set. As such, a GPR-PER model predicts zero for $n^\ddagger \in [0, \kappa], \kappa > 0$, before reviving at $\kappa$.  The use of a procedure optimising kernel noise parameters $\{\sigma, R\}$ does not change the behavior as $ n^\ddagger \to\kappa$, but does smooth the discontinuities, as illustrated in Fig.~\ref{fig:main:fig_data_gpr}(f). In letting $\kappa \gg 0$ (extremely large), we lose the uniqueness of the periodic kernel in summarising an infinite basis of oscillators, and standard Gaussian kernels (e.g. RBF, RQ) are likely to apply. 

It is possible that the choice of more complex kernels could enhance forward time series predictions via GPR, but they bring additional complications which thus far remain unresolved in relation to the current application. As one example, our ability to use numerical investigations to inform kernel design is further distorted by the need for a robust optimisation procedure, as illustrated by lower-than anticipated predictive performance observed for QPER.  Another class of GPR methods, namely, spectral mixture kernels and sparse spectrum approximation using GPR have been explored in \cite{wilson2013, quia2010}. However, these techniques also require efficient optimisation procedures to learn many unknown kernel parameters, whereas the sine-squared exponential in the periodic kernel is parameterised only by two hyper-parameters. Aside from spectral methods, the generalisation of MAT32 to higher $q + 1/2$ models probes only a subset of all possible AR($q$) processes, as the restrictions on autoregressive coefficients in Matern kernels are greater than the general case considered under an AKF in this manuscript. A detailed investigation of the application of such methods for forward prediction beyond pattern recognition and with limited computational resources, remains an area of future investigation.

\section{Conclusion \label{sec:main:Conclusion}}

In this manuscript, we provided a detailed survey of machine learning and filtering techniques applied to the problem of tracking the state of a qubit undergoing non-Markovian dephasing via a record of projective measurements.  We specifically considered the task of performing predictive estimation: learning dynamics of the system from the measurement record and then predicting evolution forward in time. To accommodate stochastic dynamics under arbitrary dephasing, and without an a priori dynamical model, we chose two Bayesian learning protocols - Gaussian Process Regression (GPR) and Kalman Filtering (KF).  All Kalman algorithms predicted the qubit state forward in time better than predicting mean qubit behaviour, indicating successful prediction, though an autoregressive approach to building the Kalman dynamical model demonstrated enhanced robustness relative to Fourier-domain approaches.  Forward prediction horizons could be arbitrarily increased for all Kalman algorithms by oversampling the underlying dephasing noise.  Our investigations included studies of both linear and non-linear measurement routines and validate the utility of the Kalman filtering framework for both.  In contrast, under GPR, we found numerical evidence that this approach enables retrodiction but not forward predictions beyond the measurement record.  

There are exciting opportunities for machine learning algorithms to increase our understanding of dynamically evolving quantum systems in real time using projective measurements. Quantum systems coupled to classical spatially or temporally varying fields may benefit from classical algorithms to analyse correlation information and enable predictive control of qubits for applications in quantum information, sensing, and the like. Moving beyond a single qubit, we anticipate that measurement records will grow in complexity allowing us to exploit the natural scalability offered by machine learning for mining large datasets. In realistic laboratory environments, the success of algorithmic approaches will be contingent on robust and computationally efficient algorithmic optimisation procedures as well as the extensions beyond Markovian dynamics studied here. The pursuit of these opportunities is the subject of ongoing research.

\section{Acknowledgments}
The LSF filter is written by V. Frey and S. Mavadia \cite{mavadia2017}. The GPR framework is implemented and optimised using standard protocols in GPy \cite{gpy2014}. Authors thank C. Granade, K. Das, V. Frey, S. Mavadia, H. Ball, C. Ferrie and T. Scholten for useful comments. This work partially supported by the ARC Centre of Excellence for Engineered Quantum Systems CE110001013, the US Army Research Office under Contract W911NF-12-R-0012, and a private grant from H. \& A. Harley.

\appendix 
\clearpage
\begin{widetext}

\section{Physical Setting \label{sec:app:setup_1}}

In this Appendix, we derive Eq.~(\ref {eqn:main:likelihood}). We consider a qubit under environmental dephasing.  For any two level system, a quantum mechanical description of physical quantities of interest can be provided in terms of the Pauli spin operators $\{ \p{x}, \p{y}, \p{z}\}$. If $\hbar \omega_A$ corresponds to an energy difference separating these two qubit states, then the Hamiltonian for a single qubit in free evolution can be written in the Pauli representation. We consider a qubit states in the $\p{z}$ basis, $\ket{0}$ or $\ket{1}$ with energies $E_0, E_1$ in our notation, corresponding to a 0 or 1 outcome upon measurement. This yields a Hamiltonian for a single qubit as:

\begin{align}
\p{z} &\equiv \ket{1}\bra{1} - \ket{0}\bra{0} \\
\op{\mathcal{I}} & \equiv \ket{0}\bra{0} + \ket{1}\bra{1} \\
\quad E_{0,1} &\equiv \mp \frac{1}{2} \hbar \omega_A \\
\op{\mathcal{H}}_0 & = \frac{1}{2} (E_0\ket{0}\bra{0}+ E_1\ket{1}\bra{1}) \\
& + \frac{1}{2} [(E_1 - E_0)\p{z} + E_0 \ket{1}\bra{1} + E_1 \ket{1}\bra{1}]\\
 &= \frac{1}{2} \hbar \omega_A \p{z}
\end{align}

In this representation, the effect of dephasing noise on a free qubit system is that any initially prepared qubit superposition of $\ket{0}$ and $\ket{1}$ states will decohere over time in the presence of dephasing noise. This physical effect is modelled as a stochastically fluctuating process $\delta\omega(t)$ that couples with the $\p{z}$ operator. The noise Hamiltonian is described as:
\begin{align} 
\op{\mathcal{H}}_{N}(t) & \equiv \frac{\hbar}{2}\delta\omega(t)\p{z}
\end{align}
In the formula above, $\delta\omega(t)$ is a classical, stochastically fluctuating parameter that models environmental dephasing and $\hbar/2$ appears as a convenient scaling factor. The total Hamiltonian for a single qubit under dephasing is:
\begin{align} 
\op{\mathcal{H}}(t) &\equiv \op{\mathcal{H}}_0 + \op{\mathcal{H}}_{N}(t)
\end{align}

Since $\op{\mathcal{H}}_{N}(t)$ commutes with $\op{\mathcal{H}}_0$, we can transform away $\op{\mathcal{H}}_0$ by moving to a rotating frame with respect to $H_0$. Let $\ket{\psi (t)}$ be a state in the lab frame, let $\op{U}$ define a transformation to a rotating frame, and let $\ket{\tilde{\psi} (t)}$ be the state in the rotating frame. The notation, $\tilde{}$, indicates operators and states in the transformed frame. In this simple case, the transformed Hamiltonian governing the evolution of $\ket{\tilde{\psi} (t)}$ will just be $\op{\mathcal{H}}_{N}(t)$:

\begin{align}
\op{U} &\equiv e^{-i\op{\mathcal{H}}_0 t / \hbar } \\
\ket{\tilde{\psi} (t)} &\equiv \op{U}^\dagger \ket{\psi (t)} \\
i \hbar \frac{d}{dt} \ket{\tilde{\psi} (t)} & \equiv i \hbar \frac{d}{dt} \op{U}^\dagger \ket{\psi (t)} \\
&= -\op{\mathcal{H}}_0 \op{U}^\dagger \ket{\psi (t)} + i\hbar \op{U}^\dagger \frac{d}{dt} \ket{\psi (t)} \\
& =  ( \op{U}^\dagger \mathcal{H}(t) \op{U} -\op{\mathcal{H}}_0  ) \ket{\tilde{\psi} (t)} \\
\implies \op{\tilde{\mathcal{H}}} &\equiv \op{U}^\dagger \mathcal{H}(t) \op{U} -\op{\mathcal{H}}_0 \\
& = \op{U}^\dagger \op{\mathcal{H}}_0\op{U}  + \op{U}^\dagger \op{\mathcal{H}}_{N}(t) \op{U} -\op{\mathcal{H}}_0 \\
& = \op{\mathcal{H}}_{N}(t),  \quad [\op{U}, \op{\mathcal{H}}_0 ] = [\op{U}, \op{\mathcal{H}}_{N}(t) ] = 0 \\
\end{align}
In the semiclassical approximation,  $\op{\mathcal{H}}_{N}(t)$ commutes with itself at different $t$, and hence we can write a unitary time evolution operator in the rotating frame as:
\begin{align}
\op{\tilde{U}}(t, t + \tau) &\equiv  e^{-\frac{i}{\hbar}  \int_{t}^{t + \tau} \op{\mathcal{H}}_{N}(t') dt'  } = e^{-\frac{i}{2} \state(t, t + \tau) \p{z} } \\
\state(t, t + \tau) & \equiv  \int_{t}^{t + \tau} \delta \omega (t') dt' \label{eqn:app:phases}
\end{align}
In the rotating frame, we prepare an initial state that is a superposition of $\ket{0}$ and $\ket{1}$ states. This state evolves under $\op{\mathcal{H}}_{N} (t)$ during a Ramsey experiment for duration $\tau$.  Subsequently, the qubit state is rotated before a projective measurement is performed with respect to the $\p{z}$ axis i.e. the measurement action resets the qubit. 

Without loss of generality, define the initial state as  $\ket{\tilde{\psi} (0)} \equiv \frac{1}{\sqrt{2}} \ket{0} + \frac{1}{\sqrt{2}} \ket{1}$ in the rotating frame. Then, the probability of measuring the same state after time $\tau$ in a single shot measurement, $d_n$ as:
\begin{align} 
Pr(d_n = 1| f(0, \tau), \tau) & = |\bra{\tilde{\psi} (0)} \op{\tilde{U}}(0, \tau) \ket{\tilde{\psi} (0)}|^2 \label{eqn:app:likelihood} \\
Pr(d_n = 0| f(0, \tau), \tau) & \equiv 1 -Pr(d_n = 1| f(0, \tau), \tau)
\end{align}
The second $\pi/2$ control pulse rotates the state vector such that a measurement in $\p{z}$ basis is possible, and the probabilities correspond to observing the qubit in the   $\ket{1}$ state. Hence, Eq.~(\ref {eqn:app:likelihood}) defines the likelihood for single shot qubit measurement. Further, Eq.~(\ref {eqn:app:likelihood}) defines the non linear measurement action on phase noise jitter, $\state(0, \tau)$.  We impose a condition that $\state(0, \tau)/2 \leq \pi$  such that accumulated phase over $\tau$ can be inferred from a projective measurement on the $\p{z}$ axis.

\subsection{Experimentally Controlled Discretisation of Dephasing Noise \label{sec:app:exptres}} 
 In this section, we consider a sequence of Ramsey measurements. At time $t$, the Eq.~(\ref {eqn:app:likelihood}) describes the qubit measurement likelihood at one instant under dephasing noise. We assume that the dephasing noise is slowly drifting with respect to a fast measurement action on timescales of order $\tau$. In this regime, Eq.~(\ref {eqn:app:phases}) discretises the continuous time process $\delta\omega(t)$, at time $t$, for a number of $n= 0, 1, ..., N$ equally spaced measurements with $t = n \Delta t$. Performing the integral for $\tau \ll \Delta t$ and slowly drifting noise such that we substitute the following terms in Eq.~(\ref {eqn:app:phases}):
\begin{align}
\delta\bar{\omega}_n &\equiv \delta\omega(t')|_{t'=n \Delta t } \\
\state_n &\equiv \state(n\Delta t, n\Delta t + \tau) \\
&=\frac{\hbar}{2}  \int_{n\Delta t}^{n\Delta t + \tau} \delta\bar{\omega}_n dt'  = \frac{\hbar}{2}\p{z}\delta\bar{\omega}_n \tau \label{eqn:app:phases_constantdetuning}
\end{align}
In this notation, $\delta\bar{\omega}_n $ is a random variable realised at time, $t = n \Delta t$, and it remains constant over short duration of the measurement action, $\tau$.  We use the shorthand $\state_n \equiv \state(n\Delta t, n\Delta t + \tau)$ to label a sequence of stochastic, temporally correlated qubit phases $ \state \equiv \{\state_n \}$. 

Since the qubit is reset by each projective measurement at $n$, the unitary operator governing qubit evolution is also reset such that \{$\op{\tilde{U}}_n \equiv \op{\tilde{U}}(n\Delta t, n\Delta t + \tau) \}$ are a collection of $N$ unitary operators describing qubit evolution for each new Ramsey experiment. They are not to be interpreted, for example, as describing qubit free evolution without re-initialising the system. Hence, for each stochastic qubit phase $\state_n$, the true probability for observing the $\ket{1}$ in a single shot is given by substituting $\state_n $ for $ \state(0,1)$ in Eq.~(\ref {eqn:app:likelihood}).
\begin{align}
Pr(d_n=d | \state_n, \tau, n \Delta t) &= \begin{cases} \cos(\frac{\state_{n}}{2})^2 \quad \text{for $d=1$} \\   \sin(\frac{\state_{n}}{2})^2  \quad \text{for $ d=0$} \end{cases} 
\end{align}
The last line follows from the fact that total probability of the qubit occupying either state must add to unity. This yields Eq.~(\ref {eqn:main:likelihood}) in the main text.

\subsection{True Dephasing Noise Engineering \label{sec:app:truenoise}} 
In the absence of an a priori model for describing qubit dynamics under dephasing noise, we impose the following properties on a sequence of stochastic phases, $\state  \equiv \{ \state_n \}$ such that we can design meaningful predictors of qubit state dynamics. We assert that a stochastic process, $\state_n$, indexed by a set of values, $ n = 0, 1, \hdots N $ satisfies: 

\begin{align}
\ex{\state_n} &= \mu_\state \quad \forall n \label{eqn:app:f_mean} \\
\ex{\state_n^2} & < \infty \quad \forall n \label{eqn:app:f_var} \\
\ex{(\state_{n_1} - \mu)(\state_{n_2} - \mu)} &= R(\nu), \quad  \nu = |n_1-n_2|, \quad \forall n_1, n_2 \in N  \label{eqn:app:f_covar} \\
R(\nu) & \neq \sigma^2  \delta(\nu) \label{eqn:app:f_Markovian} 
\end{align}
Covariance stationarity of $\state$ is established by satisfying Eqs. ~(\ref {eqn:app:f_mean}) to ~(\ref {eqn:app:f_covar}) , namely that the mean is independent of $n$, the second moments are finite, and the covariance of any two stochastic phases at arbitrary time-steps, $n_1, n_2$, do not depend on time steps but only on the separation distance, $\nu$. The $\delta(\nu)$ in the last condition,   Eq.~(\ref {eqn:app:f_Markovian}), is the Dirac-delta function and establishes that $\state$ is not delta-correlated (white). This condition captures the slowly drifting assumption for environmental dephasing noise.

We also require that correlations in $\state$ eventually die off as $\nu \to \infty$ otherwise any sample statistics inferred from noise-corrupted measurements are not theoretically guaranteed to converge to the true moments. Let $M$ be the number of runs for an experiment with $M$ different realisations of the random process $\state$, $\mu_\state$ be the true mean, $\hat{\mu}_\state$ its estimate, $\mathcal{D}_M$ denote the dataset of $M$ experiments, and $R(\nu)$ define the correlation function for the true process, $\state$. Then mean square ergodicity states that estimators approach true moments only if the correlations die off over long temporal separations:
\begin{align}
 \lim_{M \to \infty} \frac{1}{M} \sum_{\nu=0}^{M-1} R(\nu) & = 0  \iff  \lim_{M \to \infty} \ex{(\hat{\mu}_\state  - \mu_\state )^2}_{\mathcal{D}_M} = 0 \nonumber \\
\text{for} \quad \nu &= |n_{m_1} - n_{m_2}|, \quad \forall m_1, m_2 \in M, n_{m_1}, n_{m_2} \in N  \nonumber \\
\text{with} \quad \hat{\mu}_\state  &= \frac{1}{M} \sum_{m=0}^M f_{n_m} \quad   \label{eqn:app:f_msergodic}  
\end{align}
The statement above means that a true $R(\nu)$ associated with $\state$ is bandlimited for sufficiently large (but unknown) $M$. If correlations never `die out', then any designed predictors for one realisation of dephasing noise will fail for a different realisation of the same true dephasing. For the purposes of experimental noise engineering, we satisfy the assumptions above by engineering discretised process, $\state$, as:
\begin{align}
\state_n &= \alpha \omega_0 \sum_{j=1}^{J} j F(j)\cos(\omega_j n \Delta t + \psi_j) \label{eqn:app:noiseengineering} \\
F(j) & = j^{\frac{\eta}{2}-1}  
\end{align}

As described in \cite{soare2014}, $\alpha$ is an arbitrary scaling factor, $\omega_0$ is the fundamental spacing between true adjacent discrete frequencies, such that $\omega_j = 2 \pi f_0 j =\omega_0 j, j = 1, 2, ...J$. For each frequency component, there exists a uniformly distributed random phase, $\psi_j \in [0, \pi]$. The free parameter $\eta$ allows one to specify an arbitrary shape of the true power spectral density of $\state$. In particular, the free parameters $\alpha, J, \omega_0, \eta$ are true dephasing noise parameters which any prediction algorithm cannot know beforehand.

It is straightforward to show that $\state$ is covariance stationary. To show mean square ergodicity of $\state$, one requires phases are randomly uniformly distributed over one cycle for each harmonic component of $\state$ \cite{gelb1974applied}. Subsequently, one shows that an ensemble average and a long time average of multi-component engineered $\state$ are equal. For the evaluation of the long time average, we use product-to-sum formulae and observe that the case $j\neq j'$ has a zero contribution as any finite contribution from cosine terms over a symmetric integral are reduced to zero as $N \rightarrow \infty $.  For $j = j'$, only a single cosine term survives. The surviving term depends on $\nu$ and $N$ cancels to yield a finite, non-zero contribution that matches the ensemble average.

We briefly comment that $\state$ is Gaussian by the central limit theorem in the regimes considered in this manuscript. The probability density function of a sum of random variables is a convolution of the individual probability density functions. The central limit theorem grants that each element of $\state_n$ at $n$ appears Gaussian distributed for large $J$, irrespective of the underlying properties of the constituent terms or the distribution of the phases $\psi$. Numerical analysis shows that $J > 15$ results in each $\state_n$ appearing approximately Gaussian distributed. 

There is an important difference between $\state_n$ - defined here in Appendix~\ref{sec:app:setup_1} and - and $\state_n $ in Appendices~\ref{sec:app:AKF} and\nobreakspace  \ref {sec:app:spec_methods}.  In subsequent Appendices~\ref{sec:app:AKF} and\nobreakspace  \ref {sec:app:spec_methods}, the term $\state_n $  defines the `true model' for an algorithmic representation of an arbitrary covariance stationary process - either by invoking Wold's decomposition theorem (AKF, QKF) or the spectral representation theorem (LKFFB, GPR with Periodic Kernel). This means that $\state_n $ in subsequent Appendices only approximates the true covariance stationary stochastic qubit phases, $\{f_n\}$ of the Appendix~\ref{sec:app:setup_1} in the limit where total size of available sample data increases to infinity. Our notation, $f_n$, fails to distinguish these two different interpretations as such a difference does not arise in typical applications - in our case, we have no a priori true model of describing stochastic qubit phases, and must rely on mean square approximations. Henceforth, we retain $\state_n $ to be the true model for an algorithm with an understanding that this refers to an approximate representation of an arbitrary, covariance stationary sequence of stochastic qubit phases. We reserve the use of the $\hat{f_n}$ for the state estimates and predictions that an algorithm makes having considered a single noisy measurement record.

\clearpage
\section{ Autoregressive Representation of $\state$ in AKF (and QKF) \label{sec:app:AKF}}

Our objective in this Appendix is to justify the representation of $\state_n$ assumed by the AKF. In particular, we justify any $\state_n$ drawn from any arbitrary power spectral density satisfying the properties in Appendix~\ref{sec:app:truenoise} can be approximated by a high order autoregressive process.

Such results are well known, if dispersed among standard engineering and econometrics textbooks \cite{hamilton1994time,brockwell1996introduction,west1996bayesian,harvey1990forecasting,landau1998adaptive,candy2016bayesian}. We struggled to find standard references that explicitly link high $q$ AR models in approximating arbitrary covariance stationary time series of arbitrary power spectral densities, though some general comments are made in \cite{west1996bayesian}. In the discussion below, we summarise relevant background, and link a high $q$ AR process to a theorem that guarantees arbitrary representation of zero mean covariance stationary processes, and provide explicit references for proofs out of scope of introductory remarks in this Appendix. In order to achieve this, we will consider autoregressive (AR) processes of order $q$, (AR($q$)), and  moving average processes of order, $p$ (MA($p$)). A model incorporating both types of processes is known as an ARMA($q,p$) model in our notation. 

First, we define the lag operator, $\mathcal{L}$. This operator defines a map between time series sequences and enables a compact description of ARMA processes. For an infinite time series $\{ f_n \}_{n = -\infty}^{\infty}$ and a constant scalar, $c$, the lag operator is defined by the following properties:
\begin{align}
\mathcal{L} f_n & = f_{n-1} \\
\mathcal{L}^q f_n & = f_{n-q} \\
\mathcal{L}(cf_n) & = c\mathcal{L}f_n = cf_{n-1}  \\
\mathcal{L}f_n & = c, \quad \forall n, \implies \mathcal{L}^q f_n  = c
\end{align}
Next, we define a Gaussian white noise sequence, $\xi$, under the strong condition than what is stated simply in Eq.~(\ref {eqn:app:ARMA:xi_indep}), that $\xi_{n_1}, \xi_{n_2}$ are independent $\forall n_1, n_2 $:
\begin{align}
\ex{\xi} &\equiv 0  \\
\ex{\xi_{n_1} \xi_{n_2}} &\equiv \sigma^2 \delta(n_1-n_2)\label{eqn:app:ARMA:xi_indep}   
\end{align}

With these definitions, we can define an autoregressive process and a moving average process of unity order.  Eq.~(\ref {eqn:app:ARMA:AR_1}) defines an AR($q=1$) process and dynamics of $\state_n$ are given as lagged values of the variable $\state$. The second definition in Eq.~(\ref {eqn:app:ARMA:MA_1}) depicts a MA($p = 1$) process where dynamics are given by lagged values of Gaussian white noise $\xi$. 
\begin{align}
(1 - \phi_1 \mathcal{L}) f_n  & = c + \xi_n  \label{eqn:app:ARMA:AR_1} \\
f_n & = c' + (\Psi_1 \mathcal{L} + 1)\xi_n  \label{eqn:app:ARMA:MA_1} 
\end{align}
Here, $\Psi_1, \phi_1$ are known scalars defining dynamics of $\state_n$; $w_n$ is a white noise Gaussian process, and $c, c'$ are fixed scalars. It is well known that an MA($\infty$) representation is equivalently an AR($1$) process, and the reverse relationship also applies. For example, we can re-write Eq.~(\ref {eqn:app:ARMA:AR_1}) as:
\begin{align}
\state_n & = c + \xi_n + \phi_1 \state_{n-1} \\
& = w_n + \phi_1 \state_{n-1} \\
& = w_n + \phi_1 (w_{n-1}+ \phi_1 \state_{n-2} ) \\
& \vdots \\
& = \phi_1^{n+1} F_0 + \phi_1^{n} w_{0} + \phi_1^{n-1} w_{1} + \hdots w_n \\
& = \phi_1^{n+1} F_0 + \phi_1^{n} (c + \xi_{0}) + \hdots + (c + \xi_{n}) \\
& = \phi_1^{n+1} F_0 +  c (\phi_1^{n} + \phi_1^{n-1} + \hdots + 1) + \sum_{k=0}^{n} \phi_1^k \xi_{n-k} \\
w_n & \equiv c + \xi_n \\
F_0 & \equiv \state_{n=-1} 
\end{align} In the last line (and for all subsequent analysis in this Appendix),  $k$ should only be interpreted as a index variable for compactly re-writing terms in an equation as summations. We restrict $|\phi_1| < 1$ such that $\state$ is covariance stationary \cite{hamilton1994time}. 
Under these conditions, we take the limit of $\state$ capturing an infinite past, namely, as $n \to \infty$. The initial state $F_0$ is eventually forgotten, $\phi_1^{n+1} F_0 \approx 0$ if $n$ is large and $|\phi_1| < 1$. Similarly, the terms $c (\phi_1^{n} + \phi_1^{n-1} + \hdots + 1)$  can be summarised as a geometric series in $\phi_1$. The remaining terms satisfy the definition of an MA($\infty$) process:
\begin{align}
\state_n &= c \frac{1}{1 - |\phi_1|}  +  \sum_{k=0}^{\infty} \phi_1^k \xi_{n-k}, \quad |\phi_1| < 1
\end{align}
It is straightforward to show that the reverse is true, namely, an MR($1$) is equivalent to an AR($\infty$) representation \cite{hamilton1994time}.

The consideration of an MA($\infty$) process leads us directly to Wold's decomposition for arbitrary covariance stationary processes, namely, that any covariance stationary $\state$ can be represented as:
\begin{align}
\state_n & \equiv  c' + \sum_{k=0}^{\infty} \Psi_k \mathcal{L}^k \xi_{n}  \label{eqn:app:ARMA:MAinf} \\
c' & \equiv \ex{\state_n | \state_{n-1}, \state_{n-2}, \hdots} \\
\Psi_0 & \equiv 1 \\
\sum_{k=0}^{\infty} \Psi_k^2 & < \infty
\end{align}
Eq.~(\ref {eqn:app:ARMA:MAinf}) defines an MA($\infty$) process derived previously as an AR($1$) process. This process is ergodic for Gaussian $\xi$. However, such a representation of $\state$ requires fitting data to an infinite number of parameters $\{\Psi_1, \Psi_2, \hdots \}$  and approximations must be made. 


We approximate an arbitrary covariance stationary $\state$ using finite but high order AR($q$) processes. Below we show that any finite order AR($q$) process has an MA($\infty$) representation satisfying Wold's theorem.

We define an arbitrary AR($q$) process as:
\begin{align}
\xi_n & \equiv (1 - \phi_1 \mathcal{L}  - \phi_2 \mathcal{L} ^2 - \hdots -\phi_q \mathcal{L} ^q) (\state_n - c)
\end{align}
In particular, we define $\lambda_i, i = 1, \hdots, q$ as eiqenvalues of the dynamical model, $\Phi$:
\begin{align}
\Phi &\equiv \begin{bmatrix} \phi_1 & \phi_2 & \phi_3 & \hdots & \phi_{q-1}  &\phi_q \\
1 & 0 & 0 & \hdots & 0 & 0 \\
0 & 1 & 0 & \hdots & 0 & 0 \\
0 & 0 & 1 & \hdots & 0 & 0 \\
\vdots & \vdots & \vdots & \hdots & \vdots & \vdots \\
0 & 0 & 0 & \hdots & 1 & 0 \\
 \end{bmatrix} \\
\bf{\lambda} & \equiv \begin{bmatrix} \lambda_1 \dots \lambda_q \end{bmatrix} \quad \text{s.t.} |\Phi - \lambda\mathcal{I}_q|  = 0 \\
\end{align}
We use the following result from \cite{hamilton1994time} without proof that the above implies:
\begin{align}
1 & - \phi_1 \mathcal{L}  - \phi_2 \mathcal{L}^2 - \hdots -\phi_q \mathcal{L}^q \\
&\equiv (1 - \lambda_1\mathcal{L}) \hdots (1 - \lambda_q \mathcal{L}) 
\end{align}
This yields:
\begin{align}
\xi_n & = (1 - \lambda_1 \mathcal{L}) \hdots (1 - \lambda_q \mathcal{L}) (\state_n - c) \label{eqn:app:ARMA:ar_p_1}
\end{align}
For us to invert this problem and recover an MA process, we need to show that the inverse for each $(1 - \lambda_{q'} \mathcal{L})$ term exists for $q' = 1, \hdots, q$. To do this, we start by defining the operator $\Lambda_q(\mathcal{L}) $ :
\begin{align}
\Lambda_q(\mathcal{L}) & \equiv \lim_{k\to \infty} (1 + \lambda_q \mathcal{L} + \hdots + \lambda_q^k\mathcal{L}^k) \label{eqn:app:AR_inverse_0}
\end{align}
We consider an arbitrary $q'$-th eigenvalue term in process and we multiply $\Lambda_{q'}(\mathcal{L})$ :
\begin{align}
 \Lambda_{q'}(\mathcal{L}) \xi_n &= 
 \Lambda_{q'}(\mathcal{L}) (1 - \lambda_{0} \mathcal{L}) \hdots (1 - \lambda_{q'} \mathcal{L}) \hdots(\state_n - c) \\
 & = \lim_{k\to \infty} (1 + \lambda_{q'} \mathcal{L} + \hdots + \lambda_{q'}^k\mathcal{L}^k)  (1 - \lambda_{q'} \mathcal{L})(1 - \lambda_{0} \mathcal{L}) \hdots (1 - \lambda_{q'-1} \mathcal{L})  (1 - \lambda_{q'+1} \mathcal{L}) \hdots(1 - \lambda_{q} \mathcal{L})(\state_n - c) \\
 & = \lim_{k\to \infty} (1 + \lambda_{q'} \mathcal{L} + \hdots + \lambda_{q'}^k\mathcal{L}^k)(1 - \lambda_{0} \mathcal{L}) \hdots (1 - \lambda_{q'-1} \mathcal{L})  (1 - \lambda_{q'+1} \mathcal{L}) \hdots(1 - \lambda_{q} \mathcal{L})(\state_n - c) \\
  & - \lim_{k\to \infty} (\lambda_{q'} \mathcal{L} + \hdots + \lambda_{q'}^{k+1}\mathcal{L}^{k+1})(1 - \lambda_{0} \mathcal{L}) \hdots (1 - \lambda_{q'-1} \mathcal{L})  (1 - \lambda_{q'+1} \mathcal{L}) \hdots(1 - \lambda_{q} \mathcal{L})(\state_n - c) \\
& = \lim_{k\to \infty}(1 + \lambda_{q'}^{k+1}\mathcal{L}^{k+1}) (1 - \lambda_{0} \mathcal{L}) \hdots (1 - \lambda_{q'-1} \mathcal{L})  (1 - \lambda_{q'+1} \mathcal{L})\hdots (1 - \lambda_{q} \mathcal{L})(\state_n - c)
\end{align}
Each of the residual terms,  $\lambda_{q'}^{k+1}\mathcal{L}^{k+1} \to 0 $ if $|\lambda_{q'}| < 1$  for large $k$, and this case $\Lambda_{q'}(\mathcal{L})$ defines the inverse $(1 - \lambda_{q'} \mathcal{L})^{-1}$. This procedure is repeated for all $q$ eigenvalues to invert Eq.~(\ref {eqn:app:ARMA:ar_p_1}) and subsequently perform a partial fraction expansion as follows:
\begin{align}
\state_n - c & = \frac{1}{(1 - \lambda_1 \mathcal{L}) \hdots (1 - \lambda_q \mathcal{L})} \xi_n\\
& = \sum_{q'=1}^{q}\frac{a_{q'}}{1- \lambda_{q'} \mathcal{L}} \xi_n  \label{eqn:app:AR_inverse_1}\\
a_{q'} & \equiv \frac{\lambda_{q'}^{q-1}}{\prod_{q''=1, q''\neq q'}^{q} (\lambda_{q'} - \lambda_{q''})}
\end{align} The coefficients are $a_{q'}$ as obtained via the partial fraction expansion method during which $\mathcal{L}$ is treated as an ordinary polynomial. At present, we have a represent $\state$ via a finite $q$ weighted average of values of $\xi$. However, in substituting the definition of $ \Lambda_{q'} \equiv (1- \lambda_{q'} \mathcal{L})^{-1}$ from Eq.~(\ref {eqn:app:AR_inverse_0}) in Eq.~(\ref {eqn:app:AR_inverse_1}) and regrouping terms in powers of $\mathcal{L}$, we recover the form of an MA representation (setting $c \equiv \tilde{\state}_n  = 0, \quad  \forall n$ for simplicity): 
\begin{align}
\state_n & = \left[ \sum_{q'=1}^{q} a_{q'} \mathcal{L}^0 +  \lim_{k \to \infty}  \sum_{k'=1}^{k} \left( \sum_{q'=1}^{q} a_{q'}  \lambda_{q'}^{k'} \right) \mathcal{L}^{k'}\right] \xi_n \\
& = \Psi_0 + \sum_{k=1}^{\infty} \Psi_k \mathcal{L}^k \xi_{n}  \\
\Psi_0 & \equiv \sum_{q'=1}^{q} a_{q'} \mathcal{L}^0  \\
\Psi_k & \equiv \sum_{q'=1}^{q} a_{q'}  \lambda_{q'}^{k'}
\end{align}
By examining the properties of $\Phi$ raised to arbitrary powers, it can be shown that $\sum_{q'=1}^{q} a_{q'} \equiv 1$ and $\Psi_k$ is the first element of $\Phi$ raised to the $k$-the power \cite{hamilton1994time}, yielding absolute summability of $\Psi_k$ if $|\phi_{q'<q}| < 1$. This ensures that Wold's theorem is fully satisfied and an AR($p$) process has an MA($\infty$) representation. In moving to an arbitrarily high $q$, we enable the approximation of any covariance stationary $\state$.

The proofs that high $q$ AR approximations for covariance stationary $f$ improve with $q$ for example, in \cite{wahlberg1989estimation}. The key correspondence is that the number of finite lag terms $q$ in an AR($q)$) model contribute to the first $q$ values of the covariance function. This approximation improves with $q$ even if $\state$ is not a true AR process \cite{wahlberg1989estimation,west1996bayesian}. Asymptotically efficient coefficient estimates for any $MA(\infty)$ representation of $\state$ are obtained by letting the order of a purely AR($q$) process tend to infinity and increasing total data size, $N$ \cite{wahlberg1989estimation}. 

When data is fixed at $N$, we expect a high $q$ model to gradually saturate in predictive estimation performance. One can arbitrarily increase performance by increasing both $q, N$ \cite{wahlberg1989estimation}.  In our application with finite data $N$, we increase $q$ to settle on a high order AR model while training LSF to track arbitrary covariance stationary power spectral densities \cite{brockwell1996introduction}.

A high $q$ AR model is often the first step for developing models with smaller number of parameters, for example, considering a mixture of finite order AR($q$) and MA($p$) models and estimating $p+q$ number of coefficients using a range of standard protocols \cite{brockwell1996introduction,west1996bayesian}. The design of potential ARMA models for our application requires further investigation beyond the scope of this manuscript.

\clearpage
\section{Spectral Representation of $\state$ in GPR (Periodic Kernel) and LKFFB} \label{sec:app:spec_methods}

 The well-known spectral representation theorem guarantees that any covariance stationary random process (real or complex) can be represented in a generalised harmonic basis.  We defer a detailed treatment of spectral analysis of covariance stationary processes in standard textbooks, for example, \cite{hamilton1994time,karlin1975first} and present background and key results to provide insights into the choice of LKFFB and GPR (periodic kernel).
 
 The spectral representation theorem states that any covariance stationary random process has a representation given by $\state_n$, and correspondingly,  a probability distribution, $F(\omega)$ over $[-\pi, \pi]$ in the dual domain such that:
 
\begin{align}
\state_n & = \mu_\state + \int_{0}^{\pi} [ a(\omega) \cos(\omega n) +  b(\omega) \sin(\omega n) ] d\omega \\
R(\nu) & = \int_{-\pi}^{\pi} e^{-i\omega \nu } dF(\omega)
\end{align}
Here, $\mu_\state $ is the true mean of the process $\state$.  The processes $a(\omega) $ and $b(\omega)$ are zero mean and serially and mutually uncorrelated, namely, $\int_{\omega_1}^{\omega_{2}} a(\omega) d\omega$ is uncorrelated with $\int_{\omega_3}^{\omega_{4}} a(\omega) d\omega$ and $\int_{\omega_j}^{\omega_{j'}} b(\omega) d\omega$ for any $\omega_1 < \omega_2 < \omega_3 < \omega_4$ and any choice of $j, j'$ within the half cycle  $[0, \pi]$.

The distribution $F(\omega)$ exists as a limiting case of considering cumulative probability density functions for $\state_n$ at each $n$ and letting $n \to \infty$ such that a sequence of these density functions approach $F(\omega)$ \cite{karlin1975first}.  If $F(\omega)$ is differentiable with respect to $\omega$, then we see the power spectral density $S(\omega)$ and $R(\nu)$ are Fourier duals \cite{karlin1975first}:
\begin{align}
R(\nu) & = \int_{-\pi}^{\pi} e^{-i\omega \nu } S(\omega)d\omega \\
S(\omega) & \equiv \frac{dF(\omega)}{d\omega} 
\end{align}
The duality of the covariance function and the spectral density is formally expressed  in literature by the  Wiener Khinchin theorem.

We consider the finite sample analogue of the spectral representation theorem considered above by following \cite{hamilton1994time}. To proceed, we define mean square convergence as a distance metric for determining when a sequence of random variables $\{ \hat{f}_n\}$ converges to a random variable, $f_n$ in the mean square limit if:
\begin{align}
\ex{\hat{f}_n^2} & < \infty \quad \forall n \\
\lim_{n \to \infty}\ex{\hat{f}_n - f_n} & = \lim_{n \to \infty} ||\hat{f}_n- f_n || = 0
\end{align} 
The statement $||\hat{f}_n- f_n || = 0$ measures the closeness between random variables $\hat{f}_n$ and $f_n$ even though the mean square limit is defined for terms of a sequence of random variables, $\{ \hat{f}_n\}$, where convergence improves with $n \to \infty$. In context of this study, we define $\hat{f}_n$ as a linear predictor of $f_n$ belonging to a covariance stationary $\state$. Hence, each $\hat{f}_n$ for large $n$ is a linear combination of the set of random variables belonging all past noisy observations (and in Kalman Filtering,  all past state predictions). Mean square convergence of $||\hat{f}_n- f_n || = 0$ in our context is a statement of the quality of a predictor, $\hat{f}_n$ , in predicting $f_n$ as the total measurement data grows.

Next, we account for finite data and define the finite sample analogue for the spectral representation theorem. We suppose there exists a set of arbitrary, fixed frequencies  $\{\omega_j\}$  for $j = 1, \hdots , J$. We let $n$ denote finite time steps for observing $\state_n$ at $n= 1, \hdots, N$. Further, we define a set of zero mean, mutually and serially uncorrelated random process  $\{a_j \}$ and $\{b_j\}$ as finite sample analogues of the true  $a(\omega)$ and $b(\omega)$ for the $j$-th spectral component. In particular, these processes are constant over $n$ by covariance stationarity of $\state$. Then, the finite sample analogue for the spectral representation theorem becomes \cite{hamilton1994time}:
\begin{align} 
\state_n &= \mu_f  + \sum_{j=1}^{J}  [ a_j \cos(\omega_j n) +  b_j \sin(\omega_j n) ] \\
\ex{a_j} & = \ex{b_j} = 0\\
\ex{a_ja_{j'}} &= \ex{b_jb_{j'}} = \sigma^2 \delta(j - j') \\
\ex{a_jb_{j'}} &= 0 \quad \forall j, j' \\
\mu_\state &\equiv 0 
\end{align} The last line enforces a zero mean stochastic process and simplifies analysis without loss of generality  and  $\delta(\cdot)$ is the Dirac-delta function. 

To illustrate, the first two moments are of the form:
\begin{align}
\ex{f_n} &=  \mu_\state +  \sum_{j=0}^{J} E[a_j] \cos(\omega_j n) + E[b_j] \sin(\omega_j n)  = 0\\
R(\nu) &= \sum_j^{J} \sum_{j'}^{J} \sigma_j^2\delta{j,j'} [\cos(\omega_j n)\cos(\omega_j' (n+\nu)) + \sin(\omega_j n)\sin(\omega_j' (n+\nu)) ]\\
&= \sigma^2 \sum_j^{J}  p_j \cos(\omega_j\nu) \label{eqn:app:spectral:Rvtrue} \\
p_j & \equiv \frac{\sigma_j^2}{\sigma^2} \equiv \frac{\sigma_j^2}{\sum_j \sigma_j^2} 
\end{align}

We introduce process noise, $w_n$, into the formula for true $f_n$, and this establishes a commonality with state dynamics in Kalman filtering for a covariance stationary process:
\begin{align} 
\state_n &= \mu_f  + \sum_{j=1}^{J}  [ a_j \cos(\omega_j (n-1)) +  b_j \sin(\omega_j (n-1)) ] + w_n 
\end{align}

In the absence of measurement noise and operating in the oversampling regime, an ordinary least squares (OLS) regression can be constructed by providing a collection of $J^{(B)}$ basis frequencies $\{\omega_j^{(B)}\}$, as in \cite{hamilton1994time}. The OLS problem is constructed by separating the set of coefficients $\{\hat{\mu}_\state, \hat{a}_1, \hat{b}_1, \hdots \hat{a}_J, \hat{b}_J\}$ and regressors $\{1,\cos(\omega_1 (n-1)), \sin(\omega_1 (n-1)), \hdots, \cos(\omega_J^{(B)} (n-1)), \sin(\omega_J^{(B)} (n-1)) \}$. For the specific particular choice of basis,  $J^{(B)} = (N-1)/2$, (odd $N$) and $\omega_j^{(B)} \equiv 2\pi j / N$, we state the key result from \cite{hamilton1994time} that the coefficient estimates are obtained as:

\begin{align}
\hat{\state}_n &= \hat{\mu}_f  + \sum_{j=1}^{J^{(B)}}  [\hat{a}_j \cos(\omega_j^{(B)} (n-1)) +  \hat{b}_j \sin(\omega_j^{(B)} (n-1)) ] \\
\hat{a}_j &\equiv \frac{2}{N} \sum_{n'=1}^{N} \hat{\state}_{n'} \cos(\omega_j^{(B)}(n'-1)) \\
\hat{b}_j &\equiv \frac{2}{N} \sum_{n'=1}^{N} \hat{\state}_{n'} \sin(\omega_j^{(B)}(n'-1))
\end{align}
This choice of basis results in the number of regressors being the same as the length of the measurement record. Further, the term $(\hat{a}_j^2 + \hat{b}_j^2)$ is proportional to the total contribution of the $j$-th spectral component to the total sample variance of $\state$, or in other words, the amplitude estimate for the power spectral density of true $\state$.

Next, we depart from the OLS problem above by in several ways, firstly, by introducing measurement noise and secondly, by changing basis oscillators considered in the problem above. As in the main text, the linear measurement record is defined as:
\begin{align}
y_n &\equiv  f_n + v_n \\
v_n & \sim \mathcal{N}(0, R)
\end{align}
The link in GPR (periodic kernel) is direct and the link with LKFFB is made by setting $f_n \equiv H_nx_n$. In both frameworks, we incorporate the effect of measurement noise through the measurement noise variance, $R$, which has the effect of regularising the least squares estimation process discussed above.

\subsection{Infinite Basis of Oscillators in a GPR Periodic Kernel}\label{sec:ap_approxSP:GPRPKernel}

The departure from simple OLS plus measurement noise (above) to GPR (periodic kernel) arises from the fact that data is projected on an infinite basis of oscillators, namely, $J^{(B)} \to \infty$.

We follow the sketch of a proof provided in \cite{solin2014explicit} to show that a sine squared exponential (periodic kernel) used in Gaussian Process Regression satisfies covariance function of trigonometric polynomials. Here, the index $j$ labels an infinite comb of oscillators and $m$ represents the higher order terms in the power reduction formulae in the last line of the definition below:
\begin{align}
\omega_0^{(B)}  &\equiv \frac{\omega_j^{(B)} }{j}, j \in \{0, 1,..., J^{(B)}\} \\
R(\nu) &\equiv \sigma^2 \exp (- \frac{2\sin^2(\frac{\omega_0^{(B)}  \nu}{2})}{l^2}) \\
&=  \sigma^2 \exp (- \frac{1}{l^2}) \exp (\frac{\cos(\omega_0^{(B)}  \nu)}{l^2}) \label{eqn:periodic_0}\\
&=  \sigma^2 \exp (- \frac{1}{l^2}) \sum_{m = 0}^{M  \to\infty} \frac{1}{m!} \frac{\cos^m(\omega_0^{(B)}  \nu)}{l^{2m}} \label{eqn:periodic_1}
\end{align}
Next, we expand each cosine using power reduction formulae for odd and even powers respectively, and we re-group terms. For example, we expand the terms for  $m = 0,1,2,3,4,5...$ as:
\begin{align}
R(\nu) &= \sigma^2 \exp (- \frac{1}{l^2}) \cos(\omega_0^{(B)}  \nu) \left[ \frac{2}{(2l^2)}\binom{1}{0} + \frac{2}{(2l^2)^3} \frac{1}{3!} \binom{3}{1} +  \frac{2}{(2l^2)^5} \frac{1}{5!}\binom{5}{2} \dots \right] \label{eqn:cosine1}\\
& + \sigma^2 \exp (- \frac{1}{l^2}) \cos(2\omega_0^{(B)}  \nu) \left[ \frac{2}{(2l^2)^2} \frac{1}{2!} \binom{2}{0} + \frac{2}{(2l^2)^4} \frac{1}{4!} \binom{4}{1} + \dots \right] \\
& + \sigma^2 \exp (- \frac{1}{l^2}) \cos(3\omega_0^{(B)}  \nu) \left[ \frac{2}{(2l^2)^3} \frac{1}{3!} \binom{3}{0} + \frac{2}{(2l^2)^5} \frac{1}{5!}\binom{5}{1} \dots \right] \\
& + \sigma^2 \exp (- \frac{1}{l^2}) \cos(4\omega_0^{(B)}  \nu) \left[ \frac{2}{(2l^2)^4} \frac{1}{4!} \binom{4}{0} + \dots \right] \\
& + \sigma^2 \exp (- \frac{1}{l^2}) \cos(5\omega_0^{(B)}  \nu) \left[ \frac{2}{(2l^2)^5} \frac{1}{5!}\binom{5}{0} + \dots \right] \label{eqn:cosine5}\\
& \vdots \nonumber \\
& + \sigma^2 \exp (- \frac{1}{l^2}) \left[ \frac{1}{(2l^2)^2} \frac{1}{2!} \binom{2}{1} + \frac{1}{(2l)^4} \frac{1}{4!} \binom{4}{2} + \dots \right] + \sigma^2 \exp (- \frac{1}{l^2}) \label{eqn:eventerms}
\end{align}
In the expansion above, the vertical and horizontal dots represent contributions from $m>5$ terms. The key message is that truncating $m$ to a finite number of terms $M$ will   truncate $j$ to represent a finite number of oscillators. For the example above, if the power reduction expansion indexed by $m$ above was truncated to $M=5$ terms, then the  number of basis oscillators (number of rows) would also be truncated.  We now summarise the amplitudes Eq.~(\ref {eqn:cosine1}) to  Eq.~(\ref {eqn:cosine5}) in second term of $R(\nu)$ and  Eq.~(\ref {eqn:eventerms}) corresponds to $p_{0,M}$ term below:
\begin{align}
R(\nu) &= \sigma^2 (p_{0,M} + \sum_{j=0}^{\infty} p_{j,M} \cos(j\omega_0^{(B)}  \nu)) \label{eqn:app:spectral:Rvperiodic}\\
p_{j,M} & \equiv \sigma^2 \exp (- \frac{1}{l^2}) \sum_{\beta = 0}^{\beta = \beta_{j,m}^{MAX}} \frac{2}{(2l^2)^{(j + 2\beta)}} \frac{1}{(j + 2\beta)!} \binom{j + 2\beta}{\beta} \label{eqn:beta_series2} \\
\beta &\equiv  0,1,..., \beta_{j,m}^{MAX}  \\
p_{0,M} &= \exp (- \frac{1}{l^2}) \sum_{\alpha = 0}^{\alpha = \alpha_{m}^{MAX}} \frac{1}{(2l^2)^{(2\alpha)}} \frac{1}{(2\alpha)!} \binom{2\alpha}{\alpha} \label{eqn:alpha_series}\\
\alpha &\equiv  0,1,..., \alpha_{m}^{MAX} 
\end{align}
By examining the cosine expansion, one sees that a truncation at $(M, J^{(B)} )$ means our summarised formulae will require $\beta_{j,M}^{MAX} = \lfloor\frac{M-j}{2}\rfloor$ and $\alpha_{M}^{MAX} = \lfloor\frac{M}{2}\rfloor$  where $\lfloor \rfloor$ denotes the ceiling floor. If we truncate with $M \equiv J^{(B)} $ such that $\alpha_{M}^{MAX} = \lfloor\frac{J^{(B)} }{2}\rfloor, \beta_{j,M}^{MAX} =  \lfloor\frac{J^{(B)}-j}{2}\rfloor $ and re-adjust the kernel for the zero-th frequency term, then we agree with final result in \cite{solin2014explicit}.

We compare the covariance function of the periodic kernel in Eq.~(\ref {eqn:app:spectral:Rvperiodic}) with the covariance function of trigonometric polynomials in Eq.~(\ref {eqn:app:spectral:Rvtrue}).  Here, $p_{j,M}$ for the periodic kernel are not identically specified in general to those under the spectral representation theorem, but otherwise retain a structure as a cosine basis where the correlations between two random variables in a sequence only depends on the separation between them. For a constant mean Gaussian process, the form of the periodic kernel allows the underlying process to satisfy covariance stationarity and appears to permit an interpretation via the spectral representation theorem.

\subsection{Amplitude and Phase Extraction for Finite Oscillator Basis in LKFFB \label{sec:app:subsec:LKFFB}}
 In LKFFB, we depart from the simple OLS plus measurement noise problem considered earlier by specifying a fixed basis of oscillators at the physical Fourier resolution established by the measurement record. Using a specific state space model, we can track amplitudes and phases for each basis oscillator individually to enable forward prediction at any time-step of our choosing. The design of a fixed basis necessarily incorporates a priori assumptions about the extent to which a fast measurement action over-samples slowly drifting non-Markovian noise, that is, a (potentially incorrect) assumption about dephasing noise bandwidth.
 
The efficacy of the Liska Kalman Filter in our application assumes an appropriate choice of the `Kalman basis' oscillators. The choice of basis can effect the forward prediction of state estimates. To illustrate, consider the choice of Basis A - C defined below. Basis A depicts a constant spacing above the Fourier resolution (e.g. $\omega_0^{(B)} \geq \frac{2\pi}{N_T \Delta t}$). Basis B  introduces a minimum Fourier resolution and effectively creates an irregular spacing if one wishes to consider a basis frequency comb coarser than the experimentally established Fourier spacing over the course of the experiment. Basis C is identical to Basis B but allows a projection to a zero frequency component. 
 \begin{align}
 \text{Basis A: } & \equiv \{0, \omega_0^{(B)}, 2\omega_0^{(B)} \dots  J^{(B)} \omega_0^{(B)} \} \\
 \text{Basis B: } & \equiv \{ \frac{2\pi}{N \Delta t}, \frac{2\pi}{N \Delta t} + \omega_0^{(B)} , \dots,   \frac{2\pi}{N \Delta t} + J^{(B)} \omega_0^{(B)} \} \\
 \text{Basis C: } & \equiv \{ 0, \frac{2\pi}{N \Delta t}, \frac{2\pi}{N \Delta t} + \omega_0^{(B)},  \dots,   \frac{2\pi}{N \Delta t} + J^{(B)} \omega_0^{(B)} \} 
 \end{align}
 While one can propagate LKFFB with zero gain, it may be advantageous for predictive control applications to generate predictions in one calculation rather than recursively. This means we sum contributions over all $j\in J^{(B)}$ oscillators and we reconstruct the signal for all future time values in one calculation, without having to propagate the filter recursively with zero gain. The interpretation of the predicted signal, $\hat{\state}_n$, requires an additional (but time-constant) phase correction term $\psi_C$ that arises as a byproduct of the computational basis (i.e. Basis A, B or C).  The phase correction term corrects for a gradual mis-alignment between Fourier and computational grids which occurs if one specifies a non-regular spacing inherent in Basis B or C. Let $n_C$ denote the time-step at which instantaneous amplitudes $\norm{\hat{x}^j_{n_C}}$ and instantaneous phase $\theta_{\hat{x}^j_{n_C}}$ is extracted for the oscillator represented by the $j$-th state space resonator, $x^j_n $, where super-script $j$ denotes an oscillator of  frequency $\omega_j^{(B)} \equiv j\omega_0^{(B)}$ (not a power):
 \begin{align}
 \hat{f} &= \sum_{j=0}^{J^{(B)}}\norm{\hat{x}^j_{n_C}} \cos(m\Delta t \omega_j^{(B)} + \theta_{\hat{x}^j_{n_C}} + \psi_C), \\
 & \quad  n_C \in N_T, \quad m \in N_P \nonumber \\
 \psi_C & \equiv \begin{cases}
 0,  \quad \text{(Basis A)} \\
 \equiv  \frac{2\pi}{\omega_0^{(B)} } (\omega_0^{(B)} - \frac{2\pi}{N \Delta t}), \quad \text{(Basis B or C)} \\
 \end{cases}
 \end{align}
 The output predictions from calculating a harmonic sum using learned instantaneous amplitudes, phases and the LKFFB Basis A-C agree with zero-gain predictions if $\psi_C$ is specified as above. The calculation of $\psi_C$ is determined entirely by the choice of computational and experimental sampling procedures, and assumes no information about true dephasing.  
 
 Next, we define an analytical ratio to define the optimal training time, $n_C$, at which LKFFB predictions should commence, irrespective of whether the prediction procedure is recursively propagating the Kalman Filter with zero gain, or by calculating a harmonic sum for all prediction points in one go. 
 \begin{align}
 n_C &\equiv \frac{1}{\Delta t \omega_0^{(B)}} = \frac{f_s}{\omega_0^{(B)}} \label{eqn:sec:ap_liska_fixedbasis_nC}
 \end{align}
 Consider an arbitrarily chosen training period, $N_T \neq n_C $.  For $f_s$ fixed, our choice of $N_T > n_C $ means we are achieving a Fourier resolution which exceeds the resolution of the LKFFB basis. Now consider $N_T< n_C$. This means that we've extracted information prematurely, and we have not waited long enough to project on the smallest basis frequency, namely, $\omega_0^{(B)}$.  In the case where data is perfectly projected on our basis, this has no impact. For imperfect learning, we see that instantaneous amplitude and phase information slowly degrades for $N_T > n_C$; and trajectories for the smallest basis frequency have not stabilised for $N_T < n_C$. 
 
 Of these choices, Basis A for $\omega_0^{(B)} \equiv \frac{2\pi}{N_T \Delta t}$ is expected to yield best performance, at the expense of computational load, and this is confirmed in numerical experiments. All results in this manuscript are reported for Basis A with $N_T \equiv \frac{1}{\Delta t \omega_0^{(B)}} = \frac{f_s}{\omega_0^{(B)}} $.

\subsection{Equivalent Spectral Representation of $\state$ in LKFFB and GPR Periodic Kernel}

In this section, we consider the structural similarities between LKFFB and GPR with a periodic kernel. We show that the LKFFB has an analogous structure to a stack of stochastic processes on a circle \cite{karlin1975first}, and in moving from discrete to continuous time, we recover a covariance function that has the same structure if the periodic kernel was truncated to a finite basis of oscillators, $J^{(B)}$. For zero mean, Gaussian random variables, covariance stationarity is established, completing the link between LKFFB and the periodic kernel. For the case $\Gamma_n w_n \to w_n$ in LKFFB,  stacked Kalman resonators as an approximation to infinite oscillators in a periodic kernel is documented in \cite{solin2014explicit}.

At time step $n$,  the posterior Kalman state at $n-1$ acts as the initial state at $n$, such that $\nu = \Delta t$ for a small $\Delta t$ such that a linearised trajectory is approximately true for each basis frequency. We show this using the following correlation relations and a Gaussian assumption for process noise, where $n,m \in N$ are indices for time steps and $j = 0, 1, \hdots J^{(B)} $ indexes the set of basis oscillators:
 \begin{align}
 \ex{w_n} &= 0 \quad \forall j \in J^{(B)}, \quad n \in N \\
\ex{w_n,w_m} &= \sigma^2 \delta(n-m)  \quad n,m \in N \\
\ex{A^j_0} &=\ex{B^{j'}_0} = 0, \quad \forall j, j' \in J^{(B)} \\
\ex{A^j_n B^{j'}_m} &= 0, \quad \forall j, j' \in J^{(B)}, \quad n,m \in N\\
\ex{A^j_n A^{j'}_m} &= \ex{B^j_n B^{j'}_m} = \sigma_j^2 \delta(n - m)\delta(j-j'), \quad \forall j, j' \in J^{(B)}, \quad n,m \in N \\
\ex{w_n A^j_m} &= \ex{w_n B^{j'}_m} \equiv 0  \quad \forall j, j' \in J^{(B)}, \quad n,m \in N 
\end{align}
Consider a $j$-th state space resonator, $x^j_n$, in the LKFFB, where super-script $j$ denotes an oscillator (not a power) and we obtain:
\begin{align}
\Theta(j \omega_0^{(B)} \Delta t) &= \begin{bmatrix} \cos(j \omega_0^{(B)} \Delta t) & -\sin(j \omega_0^{(B)} \Delta t) \\ \sin(j \omega_0^{(B)} \Delta t) & \cos(j \omega_0^{(B)} \Delta t) \\ \end{bmatrix} \\
x^j_n & \equiv \begin{bmatrix} A^j_{n} \\ B^j_{n} \\ \end{bmatrix} = \Theta(j \omega_0^{(B)} \Delta t) \left[\idn + \frac{w_{n-1}}{\sqrt{A^j_{n-1}{}^2 + B^j_{n-1}{}^2}} \right] \begin{bmatrix} A^j_{n-1} \\ B^j_{n-1} \\ \end{bmatrix} \\
\end{align}
\begin{align}
\implies \ex{x^j_n} &= 0 \\
\implies \ex{x^j_n x^j_m{}^T}_j & =   \Theta(j \omega_0^{(B)} \Delta t) \ex{\begin{bmatrix} A^j_{n-1}A^j_{m-1} & A^j_{n-1}B^j_{m-1}\\ B^j_{n-1}A^j_{m-1} & B^j_{n-1}B^j_{m-1}\\ \end{bmatrix}} \Theta(j \omega_0^{(B)} \Delta t)^T \label{eqn:cov_kf_term1}\\
& +   \Theta(j \omega_0^{(B)} \Delta t) \left[\frac{w_{n-1}}{\sqrt{A^j_{n-1}{}^2 + B^j_{n-1}{}^2}} + \frac{w_{m-1}}{\sqrt{A^j_{m-1}{}^2 + B^j_{m-1}{}^2}} \right]\begin{bmatrix} A^j_{n-1}A^j_{m-1} & A^j_{n-1}B^j_{m-1}\\ B^j_{n-1}A^j_{m-1} & B^j_{n-1}B^j_{m-1}\\ \end{bmatrix} \Theta(j \omega_0^{(B)} \Delta t)^T  \label{eqn:cov_kf_term2}\\
& +   \Theta(j \omega_0^{(B)} \Delta t) \left[\frac{w_{n-1}w_{m-1}}{\sqrt{A^j_{n-1}{}^2 + B^j_{n-1}{}^2}\sqrt{A^j_{m-1}{}^2 + B^j_{m-1}{}^2}} \right]\begin{bmatrix} A^j_{n-1}A^j_{m-1} & A^j_{n-1}B^j_{m-1}\\ B^j_{n-1}A^j_{m-1} & B^j_{n-1}B^j_{m-1}\\ \end{bmatrix} \Theta(j \omega_0^{(B)} \Delta t)^T \label{eqn:cov_kf_term3} \\
& = \sigma^2_j \delta(n-m) \begin{bmatrix} 
1 & 0 \\ 
0 & 1  \\
\end{bmatrix} \label{eqn:cov_kf_term4}
 \end{align}

The cross correlation terms disappear under the temporal correlation functions so defined, namely, if assume $n \geq m$, then states $A^j_{m-1}, B^j_{m-1}$ at $m-1$ at most have a $w_{n-2}$ term (for the case $n=m$) and cannot be correlated with a future noise term $w_{n-1}$. 

The dynamical trajectory in LKFFB is linearised for small $\Delta t$.  The linearisation is an approximation to a true, continuous time deterministic trajectory defining a stochastic process on a circle. 

We briefly visit this continuous time trajectory to specify the link between LKFFB and GPR (periodic kernel). Let $t$ denote the continuous time deterministic dynamics for random initial state given by $a^j_0, b^j_0$, where super-script $j$ denotes an oscillator with frequency $\omega_j \equiv j \omega_0^{(B)}$ (not a power):
\begin{align}
\ex{a^j_0} &=\ex{b^{j'}_0} = 0, \quad \forall j, j' \in J^{(B)} \\
\ex{a^j_0 b^{j'}_0} &= 0, \quad \forall j, j' \in J^{(B)}\\
\ex{a^j_0 a^{j'}_0} &= \ex{b^j_0 b^{j'}_0} = \sigma_j^2\delta(j-j'), \quad \forall j, j' \in J^{(B)} \\
x^j(t) &  \equiv \begin{bmatrix} \cos(\omega_j t) & -\sin(\omega_j t) \\ \sin(\omega_j t) & \cos(\omega_j t) \\ \end{bmatrix} \begin{bmatrix} a^j_0 \\ b^j_0 \\ \end{bmatrix} \\
E[x^j(t)]&= 0 \\
E[x^j(t) x^j(t'){}^T]&= \begin{bmatrix} \cos(\omega_j t') & -\sin(\omega_j t') \\ \sin(\omega_jt') & \cos(\omega_jt') \\ \end{bmatrix} \begin{bmatrix} a^j_0 \\ b^j_0 \\ \end{bmatrix} \begin{bmatrix} a^j_0 & b^j_0 \\ \end{bmatrix} \begin{bmatrix} \cos(\omega_j t) & -\sin(\omega_j t) \\ \sin(\omega_j t) & \cos(\omega_j t) \\ \end{bmatrix} \\ 
&=\sigma_j^2 \begin{bmatrix} 
\cos(\omega_j\nu) & 0 \\ 
0 & \cos(\omega_j\nu)  \\
\end{bmatrix}, \quad \nu \equiv |t'-t| \label{eqn:cov_circle}
\end{align}
We see that the initial state variables, $a^j_0, b^j_0$, must be zero mean, independent and identically distributed variables for each $j$ such that $x^j(t)$ is covariance stationary. If $a^j_0, b^j_0$ are Gaussian, then the joint distribution, $x^j(t)$, remains Gaussian under the linear operations above. Hence, the continuous time limit of the dynamics in LKFFB for $J^{(B)}$  independent substates, $x^j(t)$, describe a process with the same first and second moments for a periodic kernel truncated at $J^{(B)}$. For Gaussian processes, this results in an approximate equivalent representation of LKFFB for $J^{(B)}$ stacked resonators with an expansion of the periodic kernel truncated at $J^{(B)}$.

While the formalism of LKFFB shares a common structure with GPR (periodic kernel) in a particular limit, the  physical interpretation of $A^j_{n}, B^j_{n}$ is that these are components of the Hilbert transform of the original signal \cite{livska2007}. This gives us the ability to track and extract instantaneous amplitude and phase associated with each basis oscillator in LKFFB. In contrast, the coefficients of the periodic kernel are always contingent on the arbitrary truncation of the infinite basis, as seen in Eqs.~(\ref {eqn:app:spectral:Rvperiodic}),  ~(\ref {eqn:beta_series2}) and ~(\ref {eqn:alpha_series}). Hence, tracking (or extracting) amplitudes and phases for individual oscillators  does not seem appropriate for the periodic kernel, as these values would change depending on the arbitrary choice of a truncation point.

\clearpage 
\clearpage
\end{widetext}




%

\end{document}